\documentclass[twocolumn]{aastex631}

\begin{document}

\title{Enhanced Pebble Drift Across Planet-Opened Gaps in Windy Protoplanetary Disks}

\author[0000-0002-0786-7307]{Lorraine Nicholson}
\affiliation{Department of Astronomy, University of Florida, Gainesville, FL 32611, USA}

\author[0000-0001-7258-770X]{Jaehan Bae}
\affiliation{Department of Astronomy, University of Florida, Gainesville, FL 32611, USA}



\begin{abstract}
When a giant planet forms in a protoplanetary disks, it carves a gap around its orbit separating the disk into two parts: inner disk and outer disk. Traditional disk accretion models, which assume material transport is driven by viscosity, reveal that the planet-induced gap acts like a filter which blocks large dust grains from flowing into the inner disk. However, there is growing evidence that material transport may be driven by magnetically-driven winds instead. By carrying out a suite of two-dimensional multi-fluid hydrodynamic simulations where wind is implemented with a parameterized model, we explored how dust filtration efficiency and the size of dust grains filtered change in disks where gas accretion is dominated by magnetically-driven winds. We found that the inward gas flow driven by the wind can enable dust to overcome the pressure bump at the outer gap edge and penetrate the planet-induced gap. The maximum size of dust grains capable of penetrating the gap increasing with the wind strength. Notably, we found that when wind is strong (mass loss rate = $10^{-7}~M_\odot~{\rm yr}^{-1}$), mm-sized grains can penetrate the gap opened by a multi-Jovian-mass planet. Our results suggest that magnetically driven winds can significantly enhance pebble drift and impact planet formation in the inner protoplanetary disk.
\end{abstract}



\section{Introduction} \label{sec:intro}
The detailed process of planet formation remains one of the great unknown frontiers in astrophysics \citep{NAS2020}. Specifically, how solid particles -- building blocks of planets -- transport and evolve to eventually form planets remain key unknowns. To understand how solid particles transport and evolve, we must examine the fundamental mechanisms driving angular momentum transport, which in turn leads to accretion. 

Traditionally, accretion in protoplanetary disks have been thought to be driven viscously. The idea behind viscous accretion is that turbulence in the disk provides viscosity, which spreads out the disk towards the central star \citep{Pringle_Rees_1972, Shakura_Sunyaev_1973}. However, our understanding of angular momentum transport in planet-forming disks has undergone a significant shift in recent years. Hydrodynamic instabilities may operate in planet-forming disks, but numerical studies showed that the level of turbulence driven by these instabilities is generally low ($\alpha \lesssim 10^{-3}$; see review by \citealt{Lesur_2023} and references therein). It has also been realized that the midplane of planet-forming disks is too weakly ionized for the magnetorotational instability \citep[MRI;][]{Balbus_Hawley_1991} to operate \citep{gammie1996}. In such environments where the gas is weakly ionized, non-ideal magnetohydrodynamic (MHD) effects become important. Indeed, MHD simulations of planet-forming disks incorporating non-ideal effects demonstrated that the MRI could be completely suppressed \citep{Bai_Stone_2013, Lesur_2014}. Instead, these simulations revealed that the disk accretion is driven by a fully laminar flow in the disk midplane, accompanied by strong magnetocentrifugal winds launched near the surface carrying away angular momentum \citep{Gressel_2015, Bai_2017, Lesur_2023}. 

Supporting these recent theoretical developments, observations with Atacama Large Millimeter/sub-millimeter Array (ALMA) suggest low levels of turbulence in protoplanetary disks. Continuum observations of inclined/edge-on protoplanetary disks revealed highly settled mm grains with a turbulent viscosity coefficient consistent with $\alpha \lesssim 10^{-3}$ \citep{pinte2016,villenave2020,villenave2022}. Moreover, molecular line observations found little to no turbulent broadening in some protoplanetary disks \citep{Flaherty_2015,Flaherty_2017,Teague_2016}.

Transport and evolution of solid particles in the traditional viscous/turbulent disk framework are relatively well studied. In a disk where the gas pressure gradually decreases with the distance from the central star, solid particles would experience aerodynamic drag and drift inward toward the star (\citealt{whipple1972}; \citealt{weidenschilling1977}). When there is an (local) inversion in the gas pressure such that the pressure increases with the distance from the star, the direction of the radial drift would be reversed and solid particles having appropriate sizes can be trapped in the pressure bump \citep{Pinilla_2012,Zhu_2012,Weber_2018,bae2019}. In fact, recent high resolution observations revealed that annular rings and gaps are common in protoplanetary disks \citep[see reviews by][]{andrews2020,bae2023}, and some of the annular rings are shown to be due to particle trapping in pressure bumps \citep{teague2018,rosotti2020}. 

However, unlike in the traditional viscous/turbulent disk framework, it has not been fully understood how solid particles transport and evolve in protoplanetary disks where accretion is governed by magnetocentrifugal winds (but see \citealt{hu2022, Wafflard_Lesur_2023}). In particular, it has not yet been investigated how the maximum size of solid particles capable of penetrating a planet-carved gap and the efficiency of dust crossing the gap change as a function of the planet mass and wind strength. To this end, in this paper we carry out a suite of two-dimensional hydrodynamic simulations where we implement magnetocentrifugal winds using a parameterized model, following \cite{Kimming_2020}. In brief, the parameterized model makes use of the magnetic lever arm and the mass loss parameter as the two parameters governing the geometry of the magnetic field and the total wind mass loss rate (see Section \ref{subsec:wind_disk} for details). This parameterization allows us to follow long-term evolution of the disk and to explore a large parameter space. 

This paper is structured as follows. In Section \ref{sec:numerical_methods} we explain the experiment design and simulation set-up including the parameterized wind-driven accretion. We present the results of our experiment in Section \ref{sec:results} and discuss implications and caveats in Section \ref{sec:Implications}. Finally, the conclusion of the study is summarized in Section \ref{sec:conclusion}.

\section{Numerical methods}\label{sec:numerical_methods}
\subsection{Hydrodynamic Equations Solved}
\label{sec:equations}

\begin{table*}[]
    \centering
    \begin{tabular}{c|c|c}
    Surface density slope   & $p$            & 0.5 \\
    Flaring index           & $(-q + 1)/2$   & 0.0 \\
    Aspect ratio at $r_p$   & $h_P$        & 0.05 \\
    Viscosity parameter     & $\alpha$     & $10^{-3}$  \\
    Smoothing length        & $s$   & 0.22$h_p$ \\
    Dust grain size         & $a$            & 1$\mu$m, 10$\mu$m, 100$\mu$m, 1mm \\    
    Planet mass  ($M_{\rm Jup}$)           & $m_p$        & 0.1, 0.3, 1, 3 \\
    Mass loss rate ($M_\odot$ yr$^{-1}$) & $\dot{M}_w$ & 0, $10^{-9}$, $10^{-8}$, $10^{-7}$ \\
    Mass loss parameter (wind) & $b$         & 0, $1.88 \times 10^{-5}$, $1.88 \times 10^{-4}$, $1.88 \times 10^{-3}$ \\
    Magnetic lever arm (wind)  & $\lambda$ & 2.25
    \end{tabular}
    \caption{Model Parameters.}
    \label{tab:model_parameters}
\end{table*}

We use the multifluid hydrodynamics code FARGO3D \citep{fargo3d, fargo_multifluid} to solve the hydrodynamic equations of motion in two-dimensional polar ($r, \phi$) coordinates, described by Equations (\ref{eqn:gas_continuity}), (\ref{eqn:dust_continuity}), (\ref{eqn:gas_momentum}), and (\ref{eqn:dust_momentum}) below. We treat dust as an inviscid and pressureless fluid, whose dynamics is described by the Euler equations. Considering a multifluid system containing gas and dust fluids, the continuity equations for the gas and dust are

\begin{equation} \label{eqn:gas_continuity}
    \frac{\partial \Sigma_g}{\partial t} + \nabla \cdot (\Sigma_g \textbf{u}) - \dot{\Sigma}_{w} = 0,
\end{equation}
\begin{equation} \label{eqn:dust_continuity}
    \frac{\partial \Sigma_d}{\partial t} + \nabla \cdot (\Sigma_d \textbf{v} + \textbf{j}) = 0,
\end{equation}
where $\Sigma_g$ and $\Sigma_d$ are the gas and dust surface density, respectively, \textbf{u} and \textbf{v} are the gas and dust velocities, and \textbf{j} denotes the mass flux due to diffusion of dust particles, for which we adopt the prescription described in detail in \citet{Weber_2018}. The term $\dot{\Sigma}_{w}$ is the mass loss rate due to wind and is further explained in Section \ref{subsec:wind_disk}.

The conservation of momentum equations for the gas and dust are
\begin{equation} \label{eqn:gas_momentum}
    \Sigma_g \left ( \frac{\partial  \textbf{u}}{\partial t} + \textbf{u} \cdot \nabla \textbf{u} \right ) = - \nabla P - \nabla \cdot \tau - \Sigma_g \nabla \Phi - \Sigma_d \textbf{f}_d - \Gamma,
\end{equation}
\begin{equation} \label{eqn:dust_momentum}
    \Sigma_d \left( \frac{\partial \textbf{v}}{\partial t} + \textbf{v} \cdot \nabla \textbf{v} \right) = - \Sigma_d \nabla \Phi + \Sigma_d \textbf{f}_d,
\end{equation}
where, $P$ is the gas pressure, $\tau$ is the viscous stress tensor, $\Phi$ is the gravitational potential, $\textbf{f}_d$ is a function representing the interaction between gas and dust species via an aerodynamic drag-force, and $\Gamma$ is the azimuthal torque due to wind (see Section \ref{subsec:wind_disk}). We assume an isothermal disk for which the pressure is related to the gas surface density and the isothermal sound speed ($c_s$) following $P = \Sigma_g c_s^2$. 

The viscous stress tensor, $\tau$, is defined as
\begin{equation} \label{eqn:viscous_stress_tensor}
    \tau \equiv \Sigma_g \nu [\nabla \textbf{u} + (\nabla \textbf{u})^T - \frac{2}{3}(\nabla \cdot \textbf{u})\textbf{I}],
\end{equation}
where \textbf{I} is the identity matrix. Throughout this study, we adopt the $\alpha$-viscosity prescription \citep{Shakura_Sunyaev_1973}, in which the viscosity is parameterized by a constant dimensionless parameter $\alpha$. We adopt $\alpha = 10^{-3}$ based on recent measurements of turbulence in protoplanetary disks \citep{Flaherty_2015, Flaherty_2017, Flaherty_2018, Flaherty_2020, Teague_2016}, which broadly suggests that a weak turbulence characterized by $\alpha \lesssim 10^{-3}$ may be common in planet-forming disks. 

The gravitational potential term, $\Phi$, includes the stellar and planet's potential: $\Phi = \Phi_* + \Phi_p$. We neglect disk's self-gravity in this study. The stellar potential $\Phi_*$ is written as 
\begin{equation} \label{eqn:star_gravPotential}
    \Phi_* = - \frac{G M_*}{r} ,
\end{equation}
where $M_*$ is the stellar mass, G is the gravitational constant, and $r$ is the distance between the star and the center of the grid cell in question. The gravitational potential of the planet is written as
\begin{equation} \label{eqn:planet_gravPotential}
    \Phi_p = - \frac{G m_p}{(|\textbf{r} - \textbf{r}_p|^2 + s^2)^{1/2}} + \frac{G m_p}{r_p^2} r \cos\phi ,
\end{equation}
where $m_p$ is the planet mass, \textbf{r} and $\textbf{r}_p$ are the radius vectors of the center of the grid cells in question and of the planet, respectively. In this study, we keep the planet on a fixed circular orbit at $r=r_p$ and ignore planetary migration to more clearly demonstrate the effect of winds in the drift/trapping of the dust. The first term of Equation (\ref{eqn:planet_gravPotential}) is the gravitational potential of the planet. The parameter $s$ in the first term is the smoothing length which is introduced to avoid an infinitely large gravitational potential at the location of the planet. For this study we adopt $s = 0.22~h_p$ (i.e., 22~\% of the local scale height), which is about the diagonal size of a grid cell at the radial location of the planet. Note that this value is smaller than what was adopted in some previous planet-disk interaction simulations where smoothing length is determined mainly to have the two-dimensional torque density matched to the three-dimensional counterpart \citep[$0.6~h_p$; e.g.,][]{Muller_2012}. In this study, we are interested in the gas and dust flows at the vicinity of the planet while fixing the planet's orbit, so we opt to use a smaller smoothing length. However, we verified that the results do not significantly change when $s = 0.6~h_p$ was used instead (see Appendix \ref{sec:smoothing_length}). The second term of Equation (\ref{eqn:planet_gravPotential}) is an indirect term which describes the acceleration experienced by the center of the frame of reference by the presence of the planet. 

For the dust fluids we consider spherical dust grains with different sizes. The dynamics of these dust fluids are governed by the Stokes number St, which is written as
\begin{equation}\label{eqn:stokes_number}
    {\rm St} \equiv \frac{\pi}{2} \frac{a \rho_{\rm int}}{\Sigma_g},
\end{equation}
where $a$ is the radius of the dust grain and $\rho_{\rm int} = 3~{\rm g~cm}^{-3}$ is the intrinsic material density of the grain. We consider 4 different sized dust grains: 1~$\mu$m, 10~$\mu$m, 100~$\mu$m, and 1~mm. With the initial gas surface density we will introduce in Section \ref{subsec:initial_conditions}, these grains initially have Stokes numbers of $2.2\times10^{-4}$, $2.2\times10^{-3}$, $2.2\times10^{-2}$, and 0.22, respectively, at the radial location of the planet.  We initialize the surface density of the total dust to be 1~\% of the gas surface density in each grid cell. We then distribute the total dust surface density to each dust species assuming the MRN grain size distribution \citep{MRN} adopting a power-law slope of $-3.5$: ${\rm d}n/{\rm d}a \propto a^{-3.5}$. Considering this grain size distribution and assuming the smallest and largest grain sizes of 0.316~$\mu$m and 3.16~mm, respectively, the fraction of mass in each dust species bin is 0.09~\%, 0.9~\%, 9~\%, and 90~\%, from smallest to largest grains. Considering the total dust-to-gas mass ratio of 0.01, the ratio between the mass in each dust species to that in the gas is $\epsilon_i = 9\times10^{-6}, 9\times10^{-5}, 9\times10^{-4},$ and $9\times10^{-3}$, respectively, from the smallest to the largest grains. 

In addition to the 4 dust species, each simulation includes one very small dust fluid having a ``fixed'' Stokes number of $10^{-6}$. Note that this dust species is treated differently from all the other dust species for which the size is fixed. Due to the small Stokes number, this small dust is well coupled to the gas in the entire simulation domain and we use this dust species to trace the gas. 

The aerodynamic drag, $\textbf{f}_d$, is described as 
\begin{equation}
    \textbf{f}_d = {\textbf{u} - \textbf{v} \over t_{\rm stop}}= {\Omega_K \over {\rm St}}(\textbf{u} - \textbf{v}).
\end{equation}
where $\Omega_K$ is the Keplerian frequency and $t_{\rm stop} \equiv {\rm St}/\Omega_K$ is the stopping time \citep{Safronov_1972}.

\subsection{Parameterized Wind Model} \label{subsec:wind_disk}
To realize MHD winds in our two-dimensional hydrodynamic simulations, we follow the methods of \cite{Kimming_2020}, which we briefly summarize below. MHD winds remove gas from the disk, which carries away angular momentum (i.e. magnetocentrifugal wind scenario; \citealt{Blandford_Payne_1982}). The lose of angular momentum drives inward accretion within the disk to conserve the overall angular momentum. In our parameterized wind model, the strength of the wind is determined by two parameters: wind loss rate $\dot{\Sigma}_{w}$ and magnetic lever arm $\lambda$. We assume that the wind loss rate is proportional to the gas surface density $\Sigma_g$ while inversely proportional to the local dynamical timescale $2\pi/\Omega_K$. Then, the wind loss rate is written as 
\begin{equation}
\label{eqn:mass_loss}
    \dot{\Sigma}_{w} = b \frac{\Omega_K}{2 \pi}{\Sigma_g},
\end{equation}
where $b$ is the mass loss parameter which determines the timescale of the mass loss.
The equation for the magnetic lever arm is 
\begin{equation} \label{lever_arm}
    \lambda = \left(\frac{r_A}{r_F}\right)^2.
\end{equation}
The lever arm describes the geometry of the magnetic field with $r_A$ being the Alfvén point and $r_F$ being the foot point of the magnetic field lines (see Figure 1 of \citealt{Kimming_2020}). Following \citet{Kimming_2020}, we adopt $\lambda = 2.25$.

The azimuthal torque associated with the mass loss in Equation (\ref{eqn:mass_loss}) can be written as
\begin{equation}
    \Gamma = \dot{\Sigma}_{w} \Omega_K r^2 (\lambda-1),
\end{equation}
which decelerates the disk gas causing inward accretion. By extracting angular momentum from the disk, the effect of the wind is that it alters the radial velocity of the gas, resulting in
\begin{eqnarray} \label{eqn:u_r_wind}
    \nonumber
    u_{r,w}  & = & -2 r \frac{\dot{\Sigma}_{w}}{\Sigma_g} (\lambda - 1)\\      & = & -\frac{b}{2\pi}(\lambda - 1) v_K,
\end{eqnarray}
where $v_K$ is the Keplerian speed. 
As a comparison, the radial gas velocity in a viscous disk is written as 
\begin{eqnarray} \label{eqn:u_r_vis}
    \nonumber
    u_{r,v} & = & -\frac{3\nu}{2r}\\
    & = & -\frac{3}{2}\alpha \left(h \over r\right)^2 v_K.
\end{eqnarray}
From Equations (\ref{eqn:u_r_wind}) and (\ref{eqn:u_r_vis}), one can see that the radial gas speed induced by the wind is equal to that in a viscous disk when 
\begin{equation} \label{eqn:b}
    b = \frac{3\pi}{\lambda - 1} \alpha \left(h \over r \right)^2.
\end{equation}
Adopting $\lambda=2.25$, $\alpha=10^{-3}$, and $h/r=0.05$, Equation (\ref{eqn:b}) tells that when $b = 1.88 \times 10^{-5}$ the wind disk model and viscous disk model have the same radial velocity. In this work, to numerically test how the strength of the wind affects the dust flow, we opt to use three $b$ values: $b = 1.88 \times 10^{-5}, 1.88 \times 10^{-4},$ and $1.88 \times 10^{-3}$. Using the initial gas surface density introduced in Section \ref{subsec:initial_conditions}, the initial wind loss rate (= accretion rate within the disk) is $\dot{M}_w = -2 \pi r \Sigma_g u_{r,w} \approx 10^{-9}, 10^{-8},$ and $10^{-7} M_\odot~{\rm yr}^{-1}$, respectively. Throughout the paper, instead of the mass loss parameter $b$ we will use the wind loss rate $\dot{M}_w$ when we refer to different models.

\subsection{Initial conditions}\label{subsec:initial_conditions}
The initial surface density and temperature profiles are described by a power-law:
\begin{equation}\label{eqn:surfaced_powerLaw}
\Sigma(r) =  \Sigma_p \left(\frac{r}{r_p}\right) ^{-p} 
\end{equation}
and
\begin{equation}\label{eqn:temperature_powerLaw}
T(r) = T_p \left( \frac{r}{r_p}\right)^{-q} , 
\end{equation}
where $\Sigma_p$ and $T_p$ are the surface density and the temperature at the location of the planet $r=1.0 r_p$. The initial gas disk mass is assumed to be $0.01 M_*$. For a system with a solar mass star and a planet located at $r_p = 30$~AU, the initial surface density $\Sigma_p = 2.11~{\rm g~cm}^{-2}$.
The disk aspect ratio corresponding to the above temperature profile is 
\begin{equation} \label{eqn:aspect_ratio}
    \frac{h}{r} = \left( \frac{h}{r}\right)_p \left(\frac{r}{r_p}\right)^{(-q + 1)/2} .
\end{equation}
Our models adopt $p = 0.5$, $q = -1.0$, and $(h/r)_P = 0.05$. 

The initial radial and azimuthal velocities are set to the steady-state analytical solution for the gas fluid,
\begin{equation}\label{eqn:u_r}
   u_r = - \frac{3 \nu}{2r} ,
\end{equation}
and
\begin{equation}\label{eqn:u_phi}
   u_\phi = v_K \sqrt{1 - \eta},
\end{equation}
where $\nu \equiv \alpha c_s^2 \Omega_K$ is the kinematic viscosity and $v_K = \Omega_K r = \sqrt{G M_*/r}$ is the Keplerian velocity. $\eta$ takes into account the additional pressure support of the disk,
\begin{equation}\label{eqn:eta}
   \eta \equiv - \left( \frac{H_g}{r}\right)^2 \frac{\partial \log P}{\partial \log r},
\end{equation}
where $H_g$ is the scale height of the gas disk which can be written as $H_g(r) = c_s(r)/\Omega_K(r)$.

We run the simulations in two stages. In the first stage, we run the simulations for 1000 planetary orbits including the gas and dust but without winds. This stage allows the gap opened by a planet to reach a quasi-steady state and ensures we have an identical starting point for the second stage where we activate winds adopting different wind strengths and introduce dust in the outer disk.

In the second stage, we use the outputs taken at the end of the first stage as the initial conditions and restart the simulations after adding the dust and activating winds. At the restart, using the same approach in \cite{Weber_2018}, the dust surface density is initialized following

\begin{equation}
\Sigma_{d,i}(r) =
\left\{
    \begin{array}{lr}
    10^{-20}~{\rm g~cm}^{-2}, & \text{if } r < 1.5 r_p\\
    \epsilon_i \Sigma_g(r), & \text{if } r \geq 1.5 r_p
    \end{array}
\right\}
\end{equation}  

where the subscript $i$ is used to indicate the $i$th dust species, $\epsilon_i$ is the ratio between the mass in the $i$th dust species to that in the gas as explained in Section \ref{sec:equations}, and $\Sigma_g(r)$ is the gas surface density at the end of the first stage. With this setup having an inner disk devoid of dust, we can ensure that the dust seen interior to the planetary orbit during the second stage is originally from the outer disk.

The dust velocity at the restart is set to the equilibrium solution \citep{Takeuchi_2002},
\begin{equation}\label{eqn:v_r}
   v_r = \frac{St^{-1} u_r - \eta v_K}{St + St^{-1}},
\end{equation}
and
\begin{equation}\label{eqn:v_phi}
   v_{\phi} = u_{\phi} - \frac{1}{2} St v_r.
\end{equation}

\begin{figure}
    \centering
    \includegraphics[scale=0.40]{ 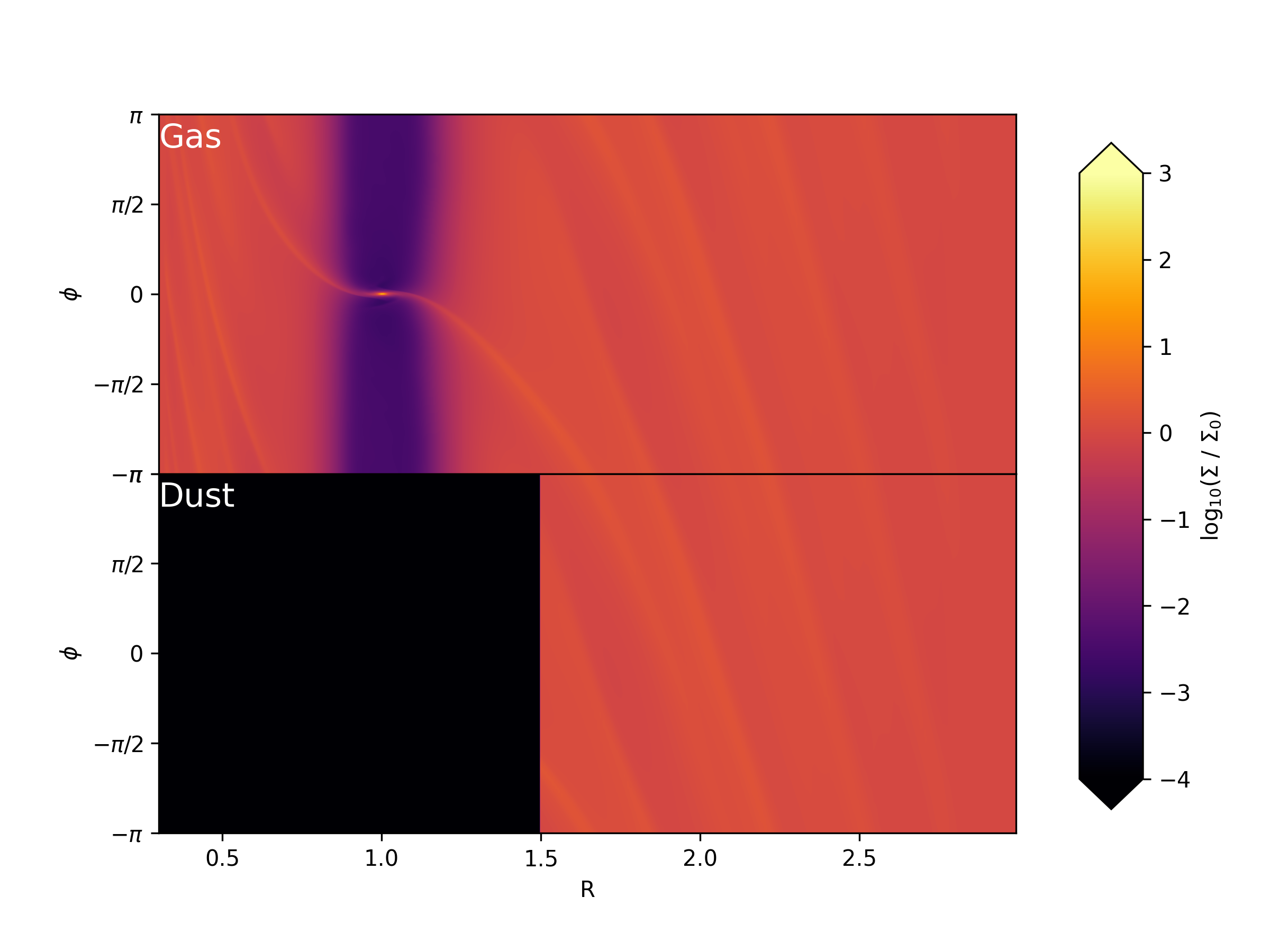}
    \caption{Example of the restart condition where dust is inserted in the outer disk only. These images show the 2D surface density distribution in polar coordinates, $(r, \phi)$. Both gas and dust surface density is normalized by the initial gas surface density, $\Sigma_{\rm g,0}$, for visualization purpose. The dust surface density is constructed such that all of the dust material is in the outer disk ($r \geq 1.5 r_p$), and in the inner disk it is equal to effectively zero. The dust surface density is scaled up from the true density according to $\epsilon_i$, so that the dust surface density in the outer disk is exactly equal to the gas surface density.}
    \label{fig:IC_image_Mjup}
\end{figure}

An example of the surface density at the start of stage two is shown in Figure. \ref{fig:IC_image_Mjup}, where we have rescaled the surface density of the dust according to $\epsilon_i$ to allow a straightforward comparison with the initial condition. With this scaling applied, the dust surface density at the beginning of stage two is identical for each dust species. Throughout the paper this upward scaling is applied to all figures regarding the surface density of dust, and the surface density is always reported in code units. 

In this work, we repeat the experiment for four cases varying the wind mass loss rate $\dot{M}_w$: no wind, 10$^{-9}$ M$_\odot$ yr$^{-1}$, 10$^{-8}$ M$_\odot$ yr$^{-1}$, and 10$^{-7}$ M$_\odot$ yr$^{-1}$. The disk mass is the same for each model $\rm M_{disk} = 0.01M_{*}$ so increasing the wind mass loss rate has the effect of shortening the disk lifetime--that is the amount of time for the gas to be completely depleted. The model with the strongest wind ($\dot{M}_w = 10^{-7} \rm ~M_\odot yr^{-1}$) has a disk lifetime of about 100,000 years, which corresponds to about 1000 orbits for a planet located at 30 au. For this reason, we chose to present the results of each model at 1000 orbits. However, we ran each of the models for 10000 orbits and verified that the result does not qualitatively change over longer timescales. The experiment was also repeated for a planet mass $m_p$ parameter study: 0.1 x Jupiter mass, 0.3 x Jupiter mass, Jupiter mass, and 3.0 x Jupiter mass. Table \ref{tab:my_label} provides a list of the models in this study with the corresponding wind mass loss rate and planet mass.

\begin{table}[]
    \centering
    \begin{tabular}{c|c|c}
    Model     &  Mass loss rate ($M_\odot~{\rm yr}^{-1}$) & Planet mass \\
    \hline
    0W1P      & 0.0 & $0.1 M_{Jup}$ \\
    1W1P      & $10^{-9}$ & $0.1 M_{Jup}$ \\
    2W1P      & $10^{-8}$ & $0.1 M_{Jup}$ \\
    3W1P      & $10^{-7}$ & $0.1 M_{Jup}$ \\
    \hline
    0W2P      & 0.0 & $0.3 M_{Jup}$ \\
    1W2P      & $10^{-9}$  & $0.3 M_{Jup}$ \\
    2W2P      & $10^{-8}$  & $0.3 M_{Jup}$ \\
    3W2P      & $10^{-7}$  & $0.3 M_{Jup}$ \\
    \hline
    0W3P      & 0.0 & $1.0 M_{Jup}$ \\
    1W3P      & $10^{-9}$  & $1.0 M_{Jup}$ \\
    2W3P      & $10^{-8}$  & $1.0 M_{Jup}$ \\
    3W3P      & $10^{-7}$  & $1.0 M_{Jup}$ \\
    \hline
    0W4P      & 0.0 & $3.0 M_{Jup}$ \\
    1W4P      & $10^{-9}$  & $3.0 M_{Jup}$ \\
    2W4P      & $10^{-8}$  & $3.0 M_{Jup}$ \\
    3W4P      & $10^{-7}$  & $3.0 M_{Jup}$ \\
    
    \end{tabular}
    \caption{List of models.}
    \label{tab:my_label}
\end{table}

\subsection{Grid Set-Up}\label{subsec:grid}
The simulation domain extends from $r_{in} = 0.3 r_p$ to $r_{out} = 3.0 r_p$ in radius and from 0 to $2 \pi$ in azimuth. For our standard run we use $N_\phi \times N_r = 1006 \times 370$ grid cells, which are linearly spaced in azimuth and logarithmically spaced in radius. The number of grid cells is chosen such that one scale height at the radial location of the planet is resolved with 8 grid cells and that grid cells at the vicinity of the planet are nearly square-shaped ($\Delta r : r \Delta \phi \simeq 1 : 1$). We ran one of our simulations with a factor of 2 lower and higher resolution and confirmed that the result does not significantly change. 

\subsection{Boundary Conditions}\label{subsec:boundary_conditions}
Boundary conditions for surface density, radial velocity, and azimuthal velocity must be determined for each fluid at the inner and outer edge of the disk. A wave damping zone (similar to \cite{deValBorro_2006}) is applied to the gas only at the inner radial boundary, and to both gas and dust at the outer radial boundary, to reduce wave reflection at the boundary and to ensure there is a steady inflow of material. The wave-damping zone takes into account the gas mass loss due to the wind so that gas is removed from the disk at a rate of $\dot{M}_w$ over time. At the inner boundary, the dust is allowed to outflow. The azimuthal velocity is set to the analytic Keplerian velocity profile. 

\section{Results}\label{sec:results}
First, we will report the results of the experiment in the case of no wind (i.e. pure viscous accretion). Then we will report the results of the experiment with wind-driven accretion and compare with the previous results. For readability we focus on the fiducial case of a Jupiter mass planet in Sections \ref{subsec:noWind} and \ref{subsec:Wind}, then summarize the results of the planet mass parameter study in Section \ref{subsec:planet_mass_param_study}. Detailed figures of the planet mass parameter study are in the Appendix A.

\begin{figure*}
    \centering
    \includegraphics[width=\textwidth]{ 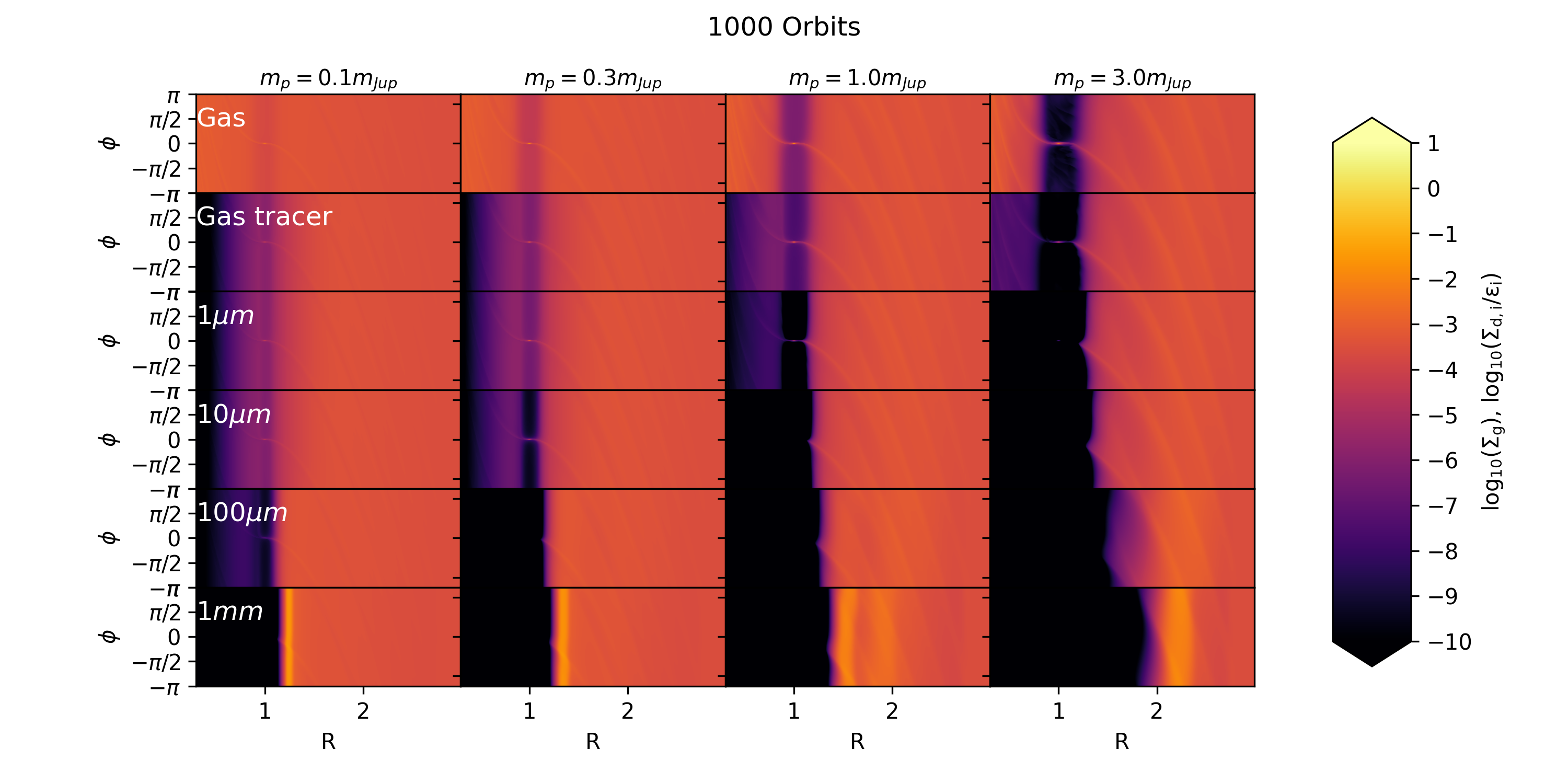}
    \caption{Two-dimensional surface density distribution in polar coordinates after 1000 orbits of evolution for models without wind. The planet it located at $(R, \phi) = (1.0, 0.0)$. The rows present different planet masses, from left to right: 0.1 M$_{Jup}$, 0.3 M$_{Jup}$, 1.0 M$_{Jup}$, 3.0 M$_{Jup}$. The columns present different fluids: from top to bottom, gas, gas tracer (St = $1 \times 10^{-6}$), 1 $\mu$m, 10 $\mu$m, 100 $\mu$m, and 1 mm. The dust surface density is scaled up from the true density according to $\epsilon_i$.}
    \label{fig:image_viscosity}
\end{figure*}
\begin{figure*}
    \centering
    \includegraphics[width=\linewidth]{ 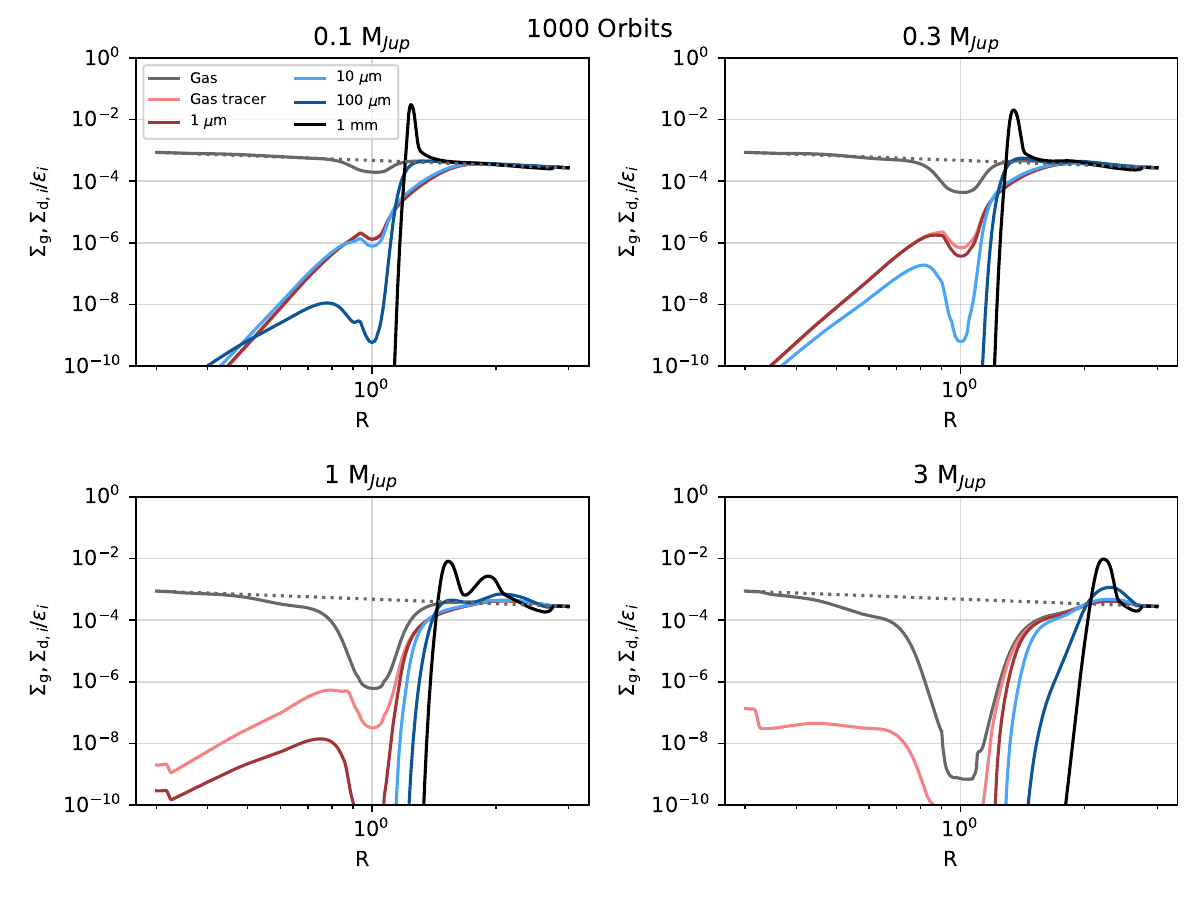}
    \caption{Azimuthally-averaged surface density profiles after 1000 orbits of evolution for models without wind, adopting various planet masses: 0.1 M$_{Jup}$ (top left), 0.3 M$_{Jup}$ (top right), 1.0 M$_{Jup}$ (bottom left), 3.0 M$_{Jup}$ (bottom right). The planet is located at $R = 1.0$. Each gas and dust fluid is plotted corresponding to the colors in the figure legend. The dotted line in each panel is the initial gas surface density according to equation \ref{eqn:surfaced_powerLaw}. The gas tracer and small dust grains were able to flow inward more efficiently than the larger dust grains; while the large dust grains are trapped in the pressure bump in the outer disk. The dust surface density is scaled up from the true density based on their abundance ($\Sigma_{d,i}/\epsilon_i$) for visualization purposes.}
    \label{fig:SD_allMp}
\end{figure*}

\subsection{Models without Wind}\label{subsec:noWind}
In the third column of Figure \ref{fig:image_viscosity}, we present the two-dimensional distribution of the gas and dust surface density for the $\rm M_p = 1.0 M_{Jup}$ model without wind, after 1000 orbits of evolution. Each row shows the result of the different dust grains, with grain size increasing from top to bottom. The gas tracer and $\rm 1 \mu m$ sized grains are seen in the inner disk, indicating that these grains were able to flow inward past the planet-induced gap. There are noticeable spiral arms which connect the material in the outer disk to the inner disk. The spiral arms provide a path for material to pass the circumplanetary disk and flow in \citep{Lubow_1999}. Note that the dust does not have time to flow all the way to the inner edge of the disk by 1000 orbits, but we verified that if we let the simulation continue longer then they will flow all the way to the edge. The other dust grains with $\rm a \geq 10 \rm \mu m$ were not able to flow into the inner disk and are trapped in the pressure bump in the outer disk because they have a larger Stokes number and are therefore decoupled from the gas. The 1 mm grains are accumulated in the outer disk which appears as bright rings around $\rm r = 1.5 r_p$ and $\rm r = 1.9 r_p$. This accumulation is coincident with a pressure maximum in the gas.

For a more quantitative comparison, we also present the azimuthally averaged surface density radial profile in Figure \ref{fig:SD_allMp} (see the bottom left panel for the $\rm M_p = M_{Jup}$ case). Again, smaller grains penetrate the gap more efficiently than larger grains because they are more coupled to the gas. Whereas larger grains do not flow inward because they are trapped in the pressure bump in the outer disk. In the case of pure viscous accretion, the planet-induced gap acts like a filter which allows small grains to flow into the inner disk but keeps large grains out. Overall, these results are broadly consistent with previous studies of dust filtration due to planet-induced gaps in viscous disks \citep[e.g.,][]{Zhu_2012,Weber_2018,bae2019}.

\begin{figure*}
    \centering
    \includegraphics[width=\textwidth]{ 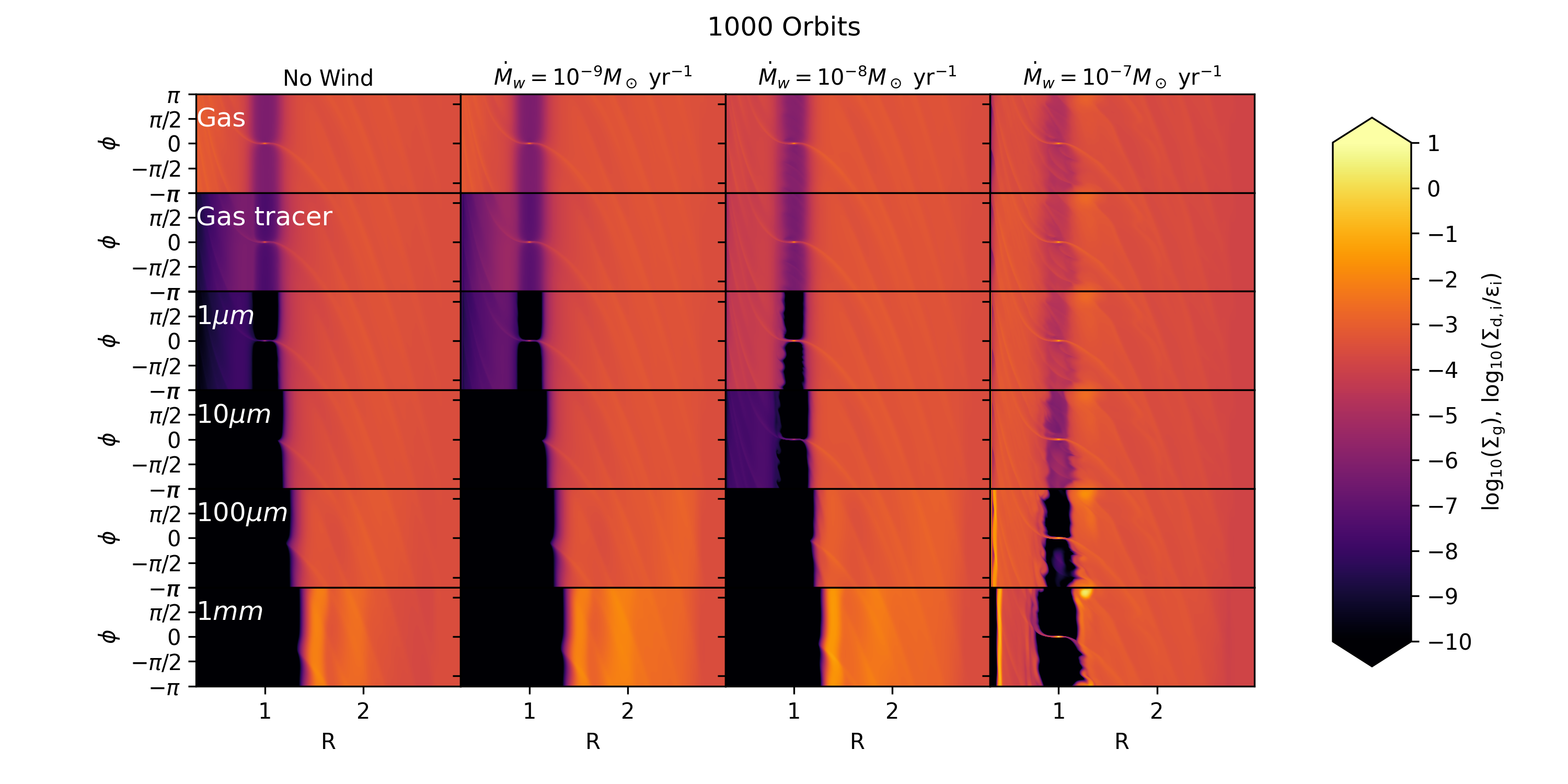}
    \caption{Two-dimensional surface density distribution in polar coordinates after 1000 orbits of evolution for a $m_p = 1 M_{\rm Jup}$ model with varying wind strength: (from left to right) no wind, $\dot{M}_w = 10^{-9} M_\odot~{\rm yr}^{-1}$, $\dot{M}_w = 10^{-8} M_\odot~{\rm yr}^{-1}$, and $\dot{M}_w = 10^{-7} M_\odot~{\rm yr}^{-1}$. The planet is located at $(R, \phi) = (1.0, 0.0)$.  The columns are different fluids in the disk, increasing from top to bottom: gas, gas tracer (St = $1 \times 10^{-6}$), 1 $\mu$m, 10 $\mu$m, 100 $\mu$m, and 1 mm. The dust surface density is scaled up from the true density according to $\epsilon_i$.}
    \label{fig:surfaceDensImage_Jup}
\end{figure*}

\subsection{Models with Wind}\label{subsec:Wind}
\begin{figure*}
    \centering
    \includegraphics[width=1.0\linewidth]{ 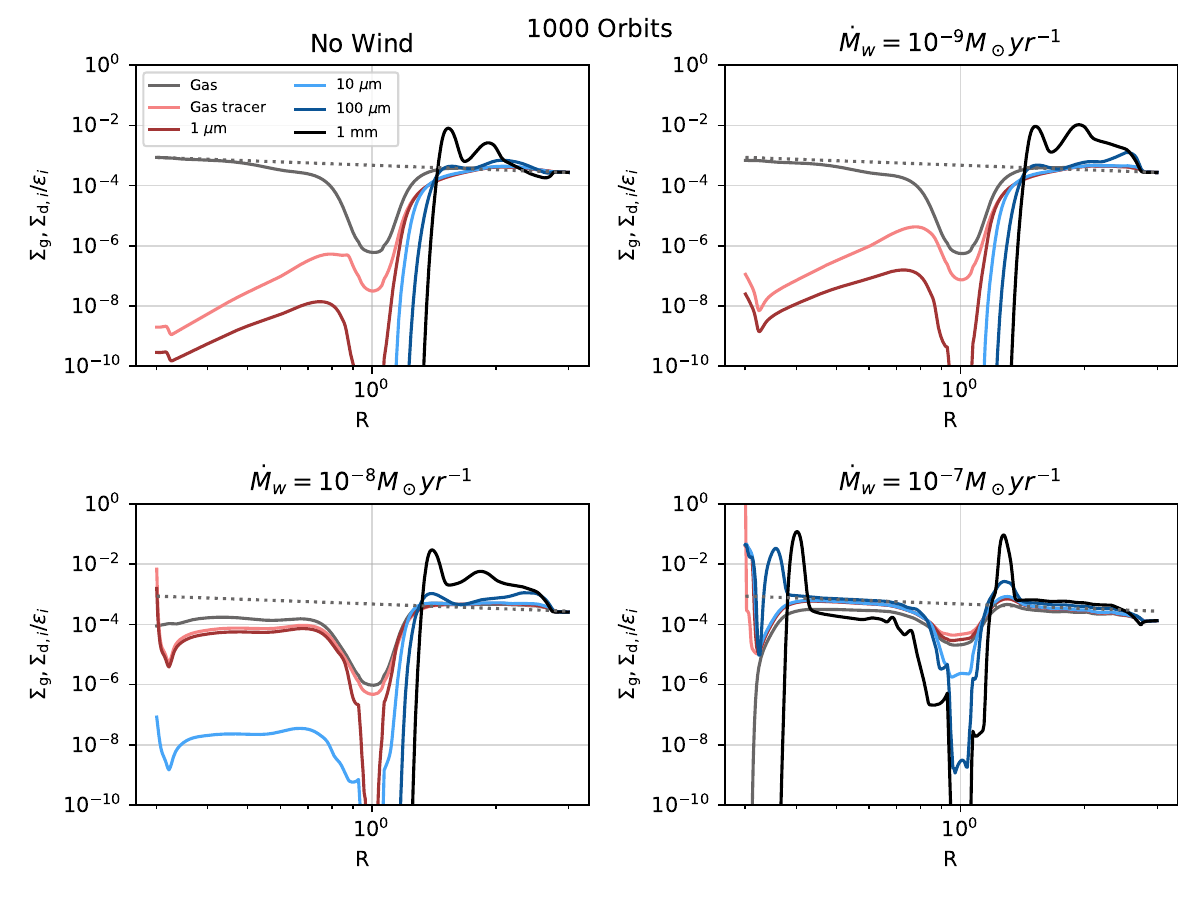}
    \caption{Azimuthally-averaged surface density profiles after 1000 orbits of evolution for models with wind, adopting various mass loss rates: No wind (top left), $\dot {M}_w = 10^{-9} M_\odot yr^{-1}$ (top right), $\dot {M}_w = 10^{-8} M_\odot yr^{-1}$ (bottom left), $\dot {M}_w = 10^{-7} M_\odot yr^{-1}$ (bottom right). The planet is located at $R = 1.0$. Each gas and dust fluid is plotted corresponding to the colors in the figure legend. The dust surface density is scaled up from the true density based on their abundance ($\Sigma_{d,i}/\epsilon_i$) for visualization purposes. The dotted line in each panel is the initial gas surface density according to equation \ref{eqn:surfaced_powerLaw}.}
    \label{fig:surfaceDens_compareBeta}
\end{figure*}

MHD-driven disk winds can potentially help dust grains flow inward and penetrate the gap more efficiently. To test this hypothesis, we repeated the above experiment while turning on the effect of disk winds (see Section \ref{subsec:wind_disk}). We considered three wind strengths corresponding to $\dot{M}_w = 10^{-9}, 10^{-8}, \text{and } 10^{-7} M_\odot~{\rm yr}^{-1}$. The resulting 2D surface density distribution after 1000 orbits of evolution for the Jupiter-mass planet case is shown in Figure \ref{fig:surfaceDensImage_Jup}. The result of the viscous model without wind is shown on the leftmost panel for comparison and once again the rows show the result for different dust grains, increasing in size from top to bottom. The complimentary azimuthally-averaged surface density profiles are shown in Figure \ref{fig:surfaceDens_compareBeta}. Overall, we found that stronger winds (i.e. higher mass loss rate $\dot{M}_w$) facilitate larger grains to penetrate the planet-opened gap. 

When $\dot{M}_w = 10^{-9} M_\odot~{\rm yr}^{-1}$ (Figure \ref{fig:surfaceDensImage_Jup} second column) there is a larger amount of 1 $\mu$m-sized grains in the inner disk then in the no wind case, indicating that even a small mass loss rate can alter the flow of material. Similar to the no wind results, the dust grains with $\rm a \geq 10 \mu m$ do not flow into the inner disk; they are trapped in the outer disk due their large Stokes number. Then, when $\dot{M}_w = 10^{-8} M_\odot~{\rm yr}^{-1}$ (Figure \ref{fig:surfaceDensImage_Jup}, third column) the effect becomes more dramatic. There is a larger amount of 1 $\mu$m-sized grains in the inner disk then in the previous case. And 10 $\mu$m-sized dust was able to flow into the inner disk. Now, only the dust grains with $\rm a \geq 100 \mu m$ are trapped in the outer disk. Finally, when $\dot{M}_w = 10^{-7} M_\odot~{\rm yr}^{-1}$ (Figure \ref{fig:surfaceDensImage_Jup}, fourth row) all four dust fluids are able to penetrate the gap and flow into the inner disk. This is because the enhanced radial gas velocity due to the wind enable large grains to overcome the pressure bump \citep{Bitsch_2018}. We will discuss this point further in Section \ref{subsec:morphology_massFlux}. 

Also, when $\dot{M}_w = 10^{-7} M_\odot~{\rm yr}^{-1}$ vortices form near the outer edge of the gap. In this case, Rossby-wave instability is triggered due to the steep surface density gradient at the gap edge which creates a vortensity minimum (similar to \citealt{Wu_2023}). Animations of Figure \ref{fig:surfaceDensImage_Jup} reveal that the vortex episodically pumps material into the inner disk whenever it passes $\phi = 0$. 

For a more quantitative comparison among our models we calculated a depletion factor, $\zeta(a)$, which quantifies how much dust filtration differs from the reference case \emph{without} a planet and \emph{without} wind. In a first attempt, we calculated $\zeta$ following the approach introduced in \cite{Weber_2018}, which is to compare the dust surface density at a single radius in the models with and without a planet. We found that this method can be applied when the wind is absent or weak ($\dot{M}_w = 10^{-9}~M_\odot~{\rm yr}^{-1}$). When the wind is strong, however, a pressure bump forms in the inner disk, trapping grains therein, and the exact radial location where the surface density peaks differ between dust species due to different Stokes number (see e.g., Figure \ref{fig:surfaceDens_compareBeta} bottom right panel). This means that the depletion factor can vary significantly depending on the exact radius it is measured. We thus opt to adopt an alternative approach, which is comparing the integrated dust mass of the inner disk: $M_d(a) = \int_{r_{in}}^{r_{out}} 2 \pi r \Sigma_d(a) dr$. With this new approach, the depletion factor is written as
\begin{equation}
    \zeta(a) = \frac{M_d (a)}{\hat{M}_d (a)},
\end{equation}
where $M_d(a)$ is the disk mass for a dust species having radius $a$ and $\hat{M}_d(a)$ is the same quantity in the reference model which has no planet and no wind. We used $r_{\rm in} = 0.35 r_p$ and $r_{\rm out} = 0.8 r_p$, although we found that $\zeta$ is not sensitive to the exact values we adopt as far as the pressure bump is located between $r_{\rm in}$ and $r_{\rm out}$.

\begin{figure*}
    \centering
    \includegraphics[width=\linewidth]{ 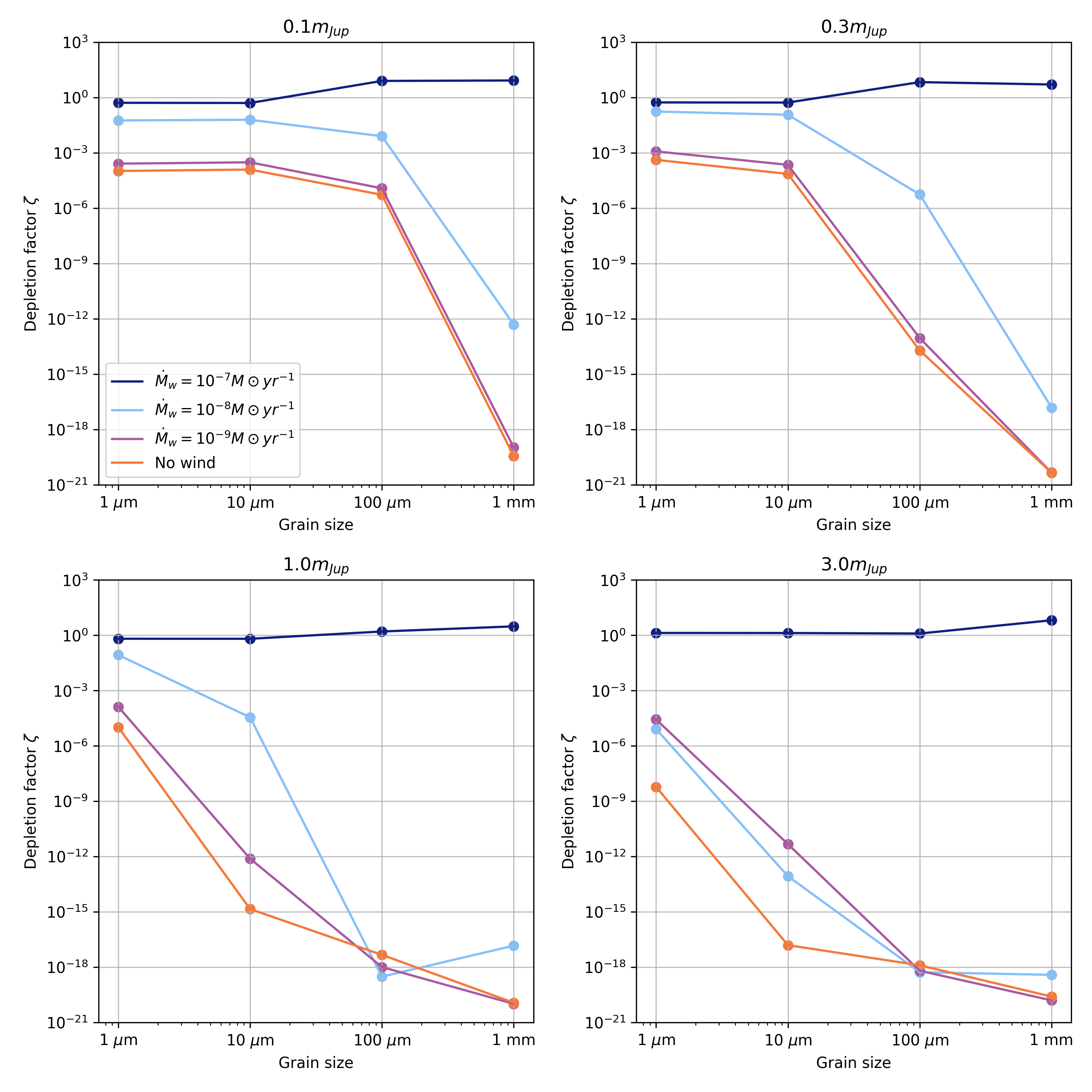}
    \caption{Depletion factor $\zeta(a)$ for every model in this study evaluated after 1000 orbits of evolution. Each panel is the result for different planet masses in this study. When there is wind, large dust grains are less depleted compared to the reference model then when there is no wind.}
    \label{fig:depletionFactor}
\end{figure*}

Figure \ref{fig:depletionFactor} shows the resulting depletion factor for each grain size for every model in this study. Let us first focus on the bottom left panel of Figure \ref{fig:depletionFactor} for the Jupiter mass planet case. When there is no wind, the depletion factor generally decreases as a function of increasing grain size. This is because when accretion is driven by viscosity, the planet-induced gap acts as a filter which does not allow large dust grains to flow into the inner disk (see Section \ref{subsec:noWind}). The smallest dust grains with $a=1.0 \rm \mu m$ are depleted by a factor of $\zeta \simeq 1.05 \times 10^{-5}$ in the inner disk compared to the reference model. Whereas, the largest dust grains with a $a = 1.0 \rm mm$ are almost completely depleted in the inner disk compared to the reference model, with the depletion factor $\zeta \simeq 0$. Overall, our results from the viscous model is in good agreement with the results from \cite{Weber_2018} where they considered pure viscous model and found that the depletion factor decreases as grain size increases. 

Now considering wind-driven accretion, when a weak wind $\dot{M}_w = 10^{-9}~M_\odot~{\rm yr}^{-1}$ is applied, the depletion factor of each grain is very similar to the case with no wind. Except for the $1.0 \rm \mu m$-sized grains which are slightly less depleted, $\zeta \simeq 1.33\times10^{-4}$, than in the no wind case. In this case, wind-driven accretion is able to facilitate small dust grains in leaking into the inner disk, but the larger dust grains are unaffected. Moving on to the intermediate wind case with $\dot{M}_w = 10^{-8}~M_\odot~{\rm yr}^{-1}$, now the smallest grains have $\zeta \simeq 0.08$ indicating that they are only slightly depleted in the inner disk compared to the reference model. The $10 \rm \mu m$-sized grains are also less depleted with $\zeta \simeq 3.56 \times 10^{-5}$ and the largest grains with $a = 1.0 \rm mm$ are depleted by a factor of $\zeta \simeq 1.47 \times 10^{-17}$, revealing that these grains are less depleted in the inner disk compared to the no wind case, despite $\zeta \ll 1$. Note that the $100 \rm \mu m$-sized grains are more depleted than the $1.0 \rm mm$-sized grains, seemingly breaking the monotonically decreasing trend. This is explained by the pressure bump in the inner disk which traps the 1mm-sized grains because of their large Stokes number (see Supplementary Figure \ref{fig:surfaceDensProfile_all}). The pressure bump forms because we remove mass from the inner edge of the disk at a rate of $\dot{M_w}$ which causes a steep gas surface density gradient at the inner edge. Finally, when there is a strong wind with $\dot{M}_w = 10^{-7}~M_\odot~{\rm yr}^{-1}$ the trend is completely reversed. Now, the depletion factor increases with increasing grain size. In particular, the 100$\rm \mu m$-sized grains have $\zeta \simeq 1.61$ and the 1.0mm-sized grains have $\zeta \simeq 3.05$, indicating that large dust grains are enhanced in the inner disk compared to the reference model. This further highlights our result that wind-dominated accretion can enhance pebble drift to the inner disk after the formation of a planet-opened gap.

\subsection{Planet mass parameter study results}\label{subsec:planet_mass_param_study}

Previous studies showed that the planet-induced gap is deeper and wider for larger planet masses in the pure viscous accretion regime\citep[e.g.,][]{Kanagawa_2015,Zhang_2018,yun2019,duffell_2020}. Figure \ref{fig:image_viscosity} shows the 2D surface density distribution after 1000 orbits of evolution for each of the planet masses in the case of no wind. And Figure \ref{fig:SD_allMp} shows the azimuthally averaged surface density profile for each of the planet masses in the case of no wind. The amount of dust material in the inner disk generally decreases with increasing planet mass. This is consistent with recent results from \cite{huang2025leakydusttrapsplanetembedded}, they also found that less massive planets create ''leakier" gaps. 

Now considering wind-driven accretion, the aforementioned trend--that stronger winds (i.e. higher mass loss rate $\dot{M}_w$) facilitate larger grains to penetrate the planet-opened gap--holds for all planet masses tested. For reference, the 2D surface density distribution after 1000 orbits for each of the planet masses are provided in the Appendix A Figures \ref{fig:surfaceDensImage_0P1Jup} to \ref{fig:surfaceDensImage_3Jup}. In all cases, when $\dot{M}_w = 10^{-9} M_\odot~{\rm yr}^{-1}$ the resulting surface density in the inner disk is similar to pure viscous accretion. That is, large dust grains are trapped in the outer disk while smaller dust grains can flow inward; and smaller planet masses allow larger grains to flow inward. On the other hand, when $\dot{M}_w = 10^{-7} M_\odot~{\rm yr}^{-1}$ then 1 mm sized dust grains were able to flow into the inner disk in every case, even when $\rm M_p = 3.0M_{Jup}$, while the surface density of these grains in the inner disk is highest for the smallest planet mass.

Similar to the Jupiter mass planet case, the depletion factor of each dust grain increases with increasing wind mass loss rate. More specifically, in the case of no wind, the smallest dust grains ($a = 1.0 \rm \mu m$) get more depleted with increasing planet mass; larger mass planets trap more small dust in the outer disk than smaller mass planets due to the gap shape. In all cases, when there is no wind or $\dot{M}_w = 10^{-9} M_\odot~{\rm yr}^{-1}$ the depletion factor decreases with increasing grain size. This is because the large grains are trapped in the outer disk. When the wind is a bit stronger ($\dot{M}_w = 10^{-8} M_\odot~{\rm yr}^{-1}$), the depletion factor of each grain is higher than the previous cases. Finally, when $\dot{M}_w = 10^{-7} M_\odot~{\rm yr}^{-1}$, the trend is reversed. Now the depletion factor increases as a function of increasing grain size. For every planet mass tested, the 1 mm-sized grains have $\zeta > 1.0$ indicating that dust is enhanced in the inner disk compared to the reference model. In conclusion, wind-dominated accretion can facilitate large dust grains to penetrate the planet-induced gap.

\subsection{Comparison with Analytic Prediction} \label{subsec:analytic_prediction}
Our simulations so far reveal a trend that increasing $\dot{M}_w$ allows larger dust grains to penetrate the gap and flow into the inner disk. This trend can be explained by the more enhanced inward radial velocity of the gas in the wind model with larger $\dot{M}_w$ (Equation \ref{eqn:u_r_wind}) which, in turn, enhance the radial velocity of the dust (Equation \ref{eqn:v_r}). We can predict the mass loss parameter b needed for a dust fluid to overcome the pressure bump in the outer disk and flow inward. Consider that the condition for dust to flow inward is simply $v_r < 0$. Using Equation (\ref{eqn:v_r}), this inequality can be expressed as

\begin{equation}
    \frac{u_{r} St^{-1} - \eta v_K}{St - St^{-1}} < 0.
\end{equation}
Now substituting Equation (\ref{eqn:u_r_wind}) for the gas radial velocity $u_r$ and rearranging to solve for $b$ results in
\begin{equation}
    b_{crit} > \frac{\pi St}{\lambda - 1} \left(\frac{c_s}{v_K}\right)^2 \frac{\partial \log P}{\partial \log r}
\end{equation}
where b$_{crit}$ is the minimum mass loss parameter required for grains with a given St to flow inward. 

\begin{figure}
    \centering
    \includegraphics[width=\linewidth]{ 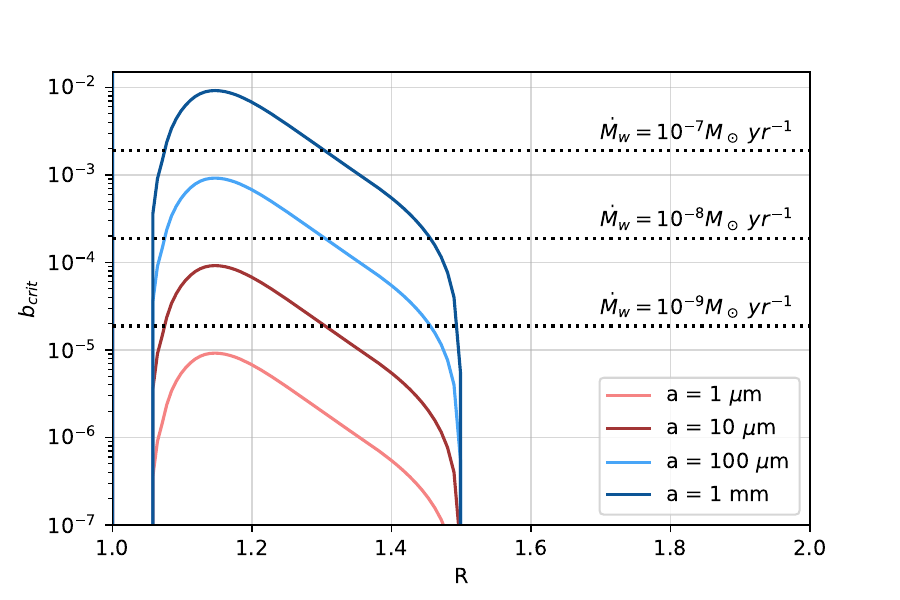}
    \caption{Analytical prediction for the mass loss parameter b$_{crit}$ that would cause a dust fluid to flow inward. Assuming $M_p = 1 M_{jup}$. For reference, we plotted b$_{crit}$ for dust fluids with a = 1 mm (blue), 100 $\mu$m (light blue), 10 $\mu$m (red), and 1 $\mu$m (pink). The dotted horizontal black lines indicate the three $\dot{M}_w$ used in this study.}
    \label{fig:b_crit}
\end{figure}

We applied this analytical prediction for the Jupiter mass planet case by numerically computing $\frac{\partial \log P}{\partial \log r}$ from the simulation output at the restart. Figure \ref{fig:b_crit} shows $b_{crit}$ as a function of radius, for various grain size values. As expected, larger dust grains require a larger critical mass loss rate in order to flow into the outer disk. In a windy disk with $\dot{M}_w = 10^{-9} \rm M_\odot ~ yr^{-1}$, as an example, grains smaller than a few $\mu$m should flow inward whereas grains larger than that should be trapped in the outer disk. For $\dot{M}_w = 10^{-8} \rm M_\odot ~ yr^{-1}$, grains smaller than a few tens $\mu$m can flow into the inner disk. Finally, we predict for $\dot{M}_w = 10^{-7} \rm M_\odot ~ yr^{-1}$ grains smaller than a few hundreds $\mu$m can flow into the inner disk and there is minimal trapping in the outer disk. This broadly matches the aforementioned trend seen in numerical simulations.

\section{Discussion}\label{sec:Implications}
\subsection{Morphology of flows and mass flux} \label{subsec:morphology_massFlux}
In this study, we found that wind-driven accretion enhances the inward dust flow. To further explore this idea we investigated the 2D velocity distribution and mass flux. This analysis was performed for the case of $M_p = 1M_{Jup}$.

\begin{figure*}
    \centering
    \includegraphics[width=\linewidth]{ 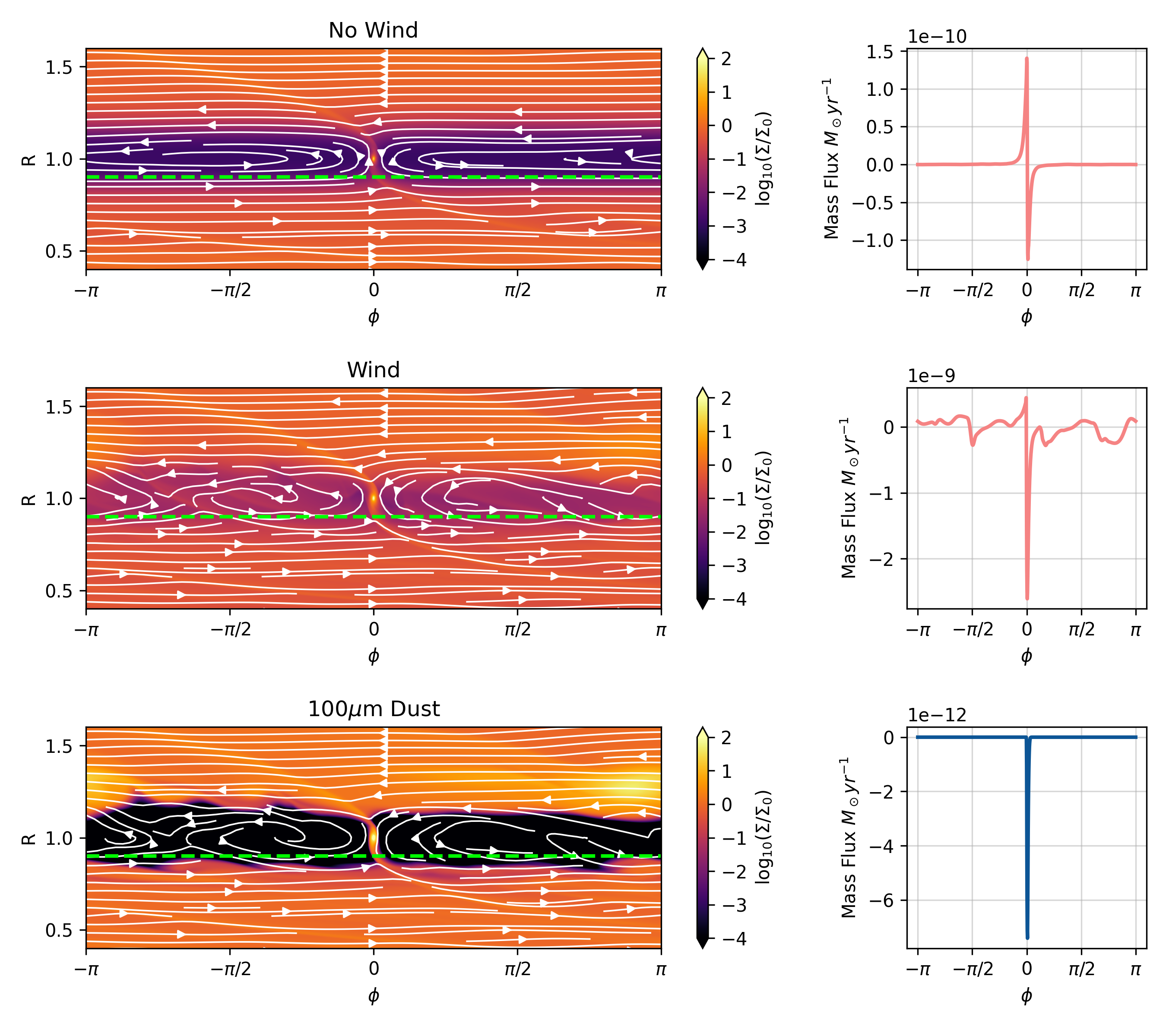}
    \caption{Left panels: Visualization of the two-dimensional velocity ($v_r, v_\phi$) distribution for the case of a Jupiter mass planet. The top panel is the gas tracer in the case of no wind, the middle panel is the gas tracer in the case of wind with mass loss rate $\dot{M}_w = 10^{-7} M_\odot~{\rm yr}^{-1}$, and the bottom panel is the 100 $\mu$m dust in the case of wind with mass loss rate $\dot{M}_w = 10^{-7} M_\odot~{\rm yr}^{-1}$. The white streamlines indicate the direction (but not magnitude) of the velocity vector. The background color contour shows the corresponding surface density. Right panels: Mass flux at each grid cell evaluated at R = 0.9 (dashed green line on the left side). Gas and dust flows in the co-rotating region near the planet become asymmetric when wind is turned-on.}
    \label{fig:massFlux_azimuth}
\end{figure*}

The top and middle panels of Figure \ref{fig:massFlux_azimuth} show the velocity vector field of the gas tracer for two cases: no wind and strong wind with $\dot{M}_w = 10^{-7} M_\odot~{\rm yr}^{-1}$. Additionally, in the right panels of Figure \ref{fig:massFlux_azimuth}, we present the mass flux at $r=0.9r_p$, computed as $\dot{M} = 2 \pi r \Sigma_r u_r$.
In the case of no wind, the mass flux spikes on either side of the planet and it is nearly symmetric on either side; in the co-orbiting region the inward flow at $\phi > \phi_p$ is comparable to the outward flow at $\phi < \phi_p$. When there is wind with $\dot{M}_w = 10^{-7} M_\odot~{\rm yr}^{-1}$, however, the picture is very different. The flow in the co-rotating region is asymmetric and the inward flux at $\phi > \phi_p$ is greater than the outward flux at $\phi < \phi_p$ by a factor of $\simeq 5.0$. On the trailing side of the planet ($\phi < \phi_p$) there is a circular recycling pattern, whereas on the leading side ($\phi > \phi_p$) there is a fast accreting flow moving towards the inner disk. Although our simulations did not include magnetic fields, the flow patterns seen in our simulations broadly resemble those seen in non-ideal MHD simulations \citep{Wafflard_Lesur_2023, Wafflard_Lesur_2025, hu_2024}. Also, the material is flowing inward faster than in the case of no wind. When there is no wind, the maximum mass flux for the gas tracer is $\simeq-1 \times 10^{-10} M_\odot~{\rm yr}^{-1}$ versus for the strong wind it is $\simeq -2 \times 10^{-9} M_\odot~{\rm yr}^{-1}$ (note that the negative sign indicates an inward flow towards the central star). 

To clarify why wind-driven accretion is able to cause large dust grains to leak into the inner disk, even after the formation of a planet opened gap, the bottom panel of Figure \ref{fig:massFlux_azimuth} shows the result for the 100 $\mu$m-sized dust. We find that the asymmetry in the velocity field of the gas tracer extends to the dust grains as well. Similar to the gas tracer, the velocity distribution in the co-rotating region is asymmetric, with more tightly packed flows pointing inward. The inward mass flux for the $100 \mu m$-sized dust reaches to $\simeq 6 \times 10^{-12} M_\odot~{\rm yr}^{-1} \approx 2 M_\oplus~{\rm Myr}^{-1}$. 

\begin{figure}
    \centering
    \includegraphics[width=1.0\linewidth]{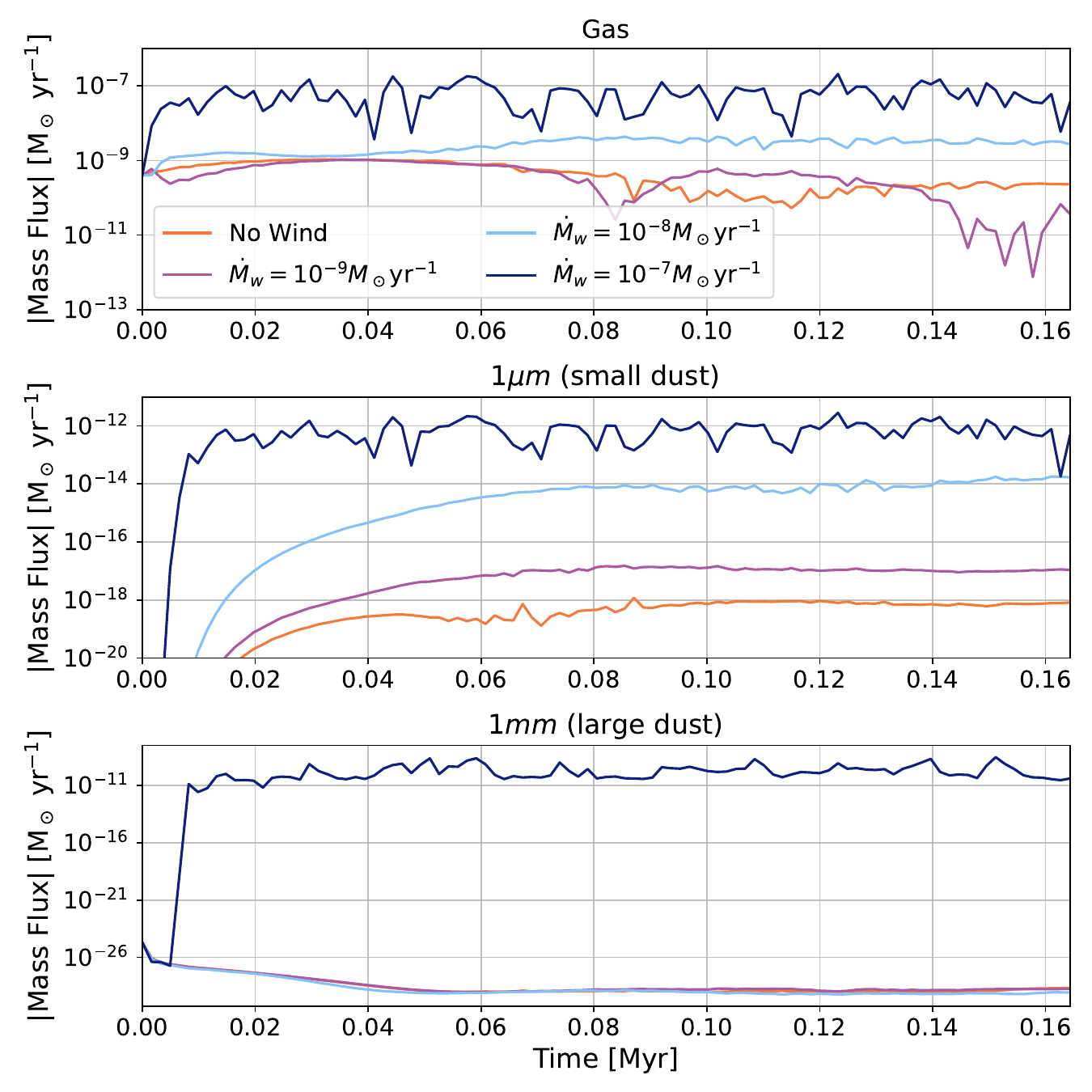}
    \caption{Absolute value of the total mass flux at $\rm R=0.9 r_p$ over time for the case of $m_p = m_{Jup}$. The different colors show different wind strengths corresponding to the figure legend. From top to bottom the panels show the result for gas, $1 \mu m$ dust, and $1 mm$ dust.}
    \label{fig:massFlux_windStrength_grainSize}
\end{figure}

In general, a sufficient pebble flow is necessary to accumulate enough dust to form a planet. Our results suggest that the mass flux of dust grains towards the inner disk can be enhanced by wind-driven accretion, which therefore has important implications for subsequent planet formation interior to a giant planet's orbit. In Figure \ref{fig:massFlux_windStrength_grainSize} we compare the absolute value of the total mass flux at $\rm R = 0.9 r_p$ over time for different wind strength. This analysis was done for the case of $\rm M_p = M_{Jup}$ (but see Figure \ref{fig:massFlux_windStrength_grainSize_all} for planet mass parameter study). Most importantly, in every panel, the mass flux is highest when $\dot{M}_w = 10^{-7} M_\odot \rm ~yr^{-1}$.

First, focusing on the mass flux of gas (Figure \ref{fig:massFlux_windStrength_grainSize} top panel), the mass flux increases for increasing wind mass loss rate. When there is no wind, the gas mass flux is fairly constant throughout the simulation with small fluctuation. Then when there is wind-driven accretion with $\dot{M}_w = 10^{-9} M_\odot \rm ~yr^{-1}$ we notice that the fluctuation is more pronounced, compared to the case of no wind, and the average value  is slightly higher. The higher mass flux indicates that wind-driven accretion enhances the inward flow of material. In the intermediate case when $\dot{M}_w = 10^{-8} M_\odot \rm ~yr^{-1}$ the average mass flux is even higher then the previous cases. Finally, in the strongest wind case when $\dot{M}_w = 10^{-7} M_\odot \rm ~yr^{-1}$, on average the gas flows in at a rate just below $10^{-7} M_\odot \rm ~yr^{-1}$. Overall, the measured gas accretion rate is broadly consistent with the wind mass loss rate adopted in each model.

Now, focusing on the mass flux of the smallest dust grains with $\rm a = 1 \mu m$ (Figure \ref{fig:massFlux_windStrength_grainSize} middle panel), the mass flux increases for increasing wind mass loss rate following the aforementioned trend in the gas. In particular, when there is no wind the small grains flow in slowly at an average rate of $10^{-18} M_\odot \rm ~yr^{-1}$. Then when there is wind-driven accretion with $\dot{M}_w = 10^{-9} M_\odot \rm ~yr^{-1}$ the mass flux is enhanced. In the intermediate wind case the mass flux is even higher. Finally, in the strong wind case with $\dot{M}_w = 10^{-7} M_\odot \rm ~yr^{-1}$ the small dust grains flow in at a rate of $10^{-12} M_\odot \rm ~yr^{-1}$ which is relatively high compared to the case of no wind.

For the largest dust grains with $a = 1\rm mm$ (Figure \ref{fig:massFlux_windStrength_grainSize} bottom panel) the picture is a little different. On one hand, for most models (no wind, $\dot{M}_w = 10^{-9} M_\odot \rm ~yr^{-1}$, and $\dot{M}_w = 10^{-8} M_\odot \rm ~yr^{-1}$) the average mass flux is $< 10^{-27} M_\odot \rm ~yr^{-1}$ which is a negligible amount of mass being transported to the inner disk. That is because in these cases, dust trapping in the outer disk is very efficient for the large grains so they do not flow inward. On the other hand, when $\dot{M}_w = 10^{-7} M_\odot \rm ~yr^{-1}$, the large dust grains flow in at a rate around $10^{-10} M_\odot \rm ~yr^{-1}$. It's possible that this efficient pebble flow, due to wind-driven accretion, may aid in subsequent planet formation in the inner disk even after the formation of a planet-opened gap however the enhancement of pebbles will be short lived due to the disk lifetime.

The fluctuations in mass flux when $\dot{M}_w = 10^{-7} M_\odot \rm ~yr^{-1}$ can be attributed to the presence of a vortex (see Figure \ref{fig:surfaceDens_compareBeta}). Over time, the vortex helps pump additional material into the inner disk episodically whenever it crosses the azimuthal location of the outer spiral arm. This effect may further enhance pebble drift into the inner disk when the wind mass loss rate is high.

\subsection{Dust-to-Gas Ratio} \label{subsec:dtg_ratio}
A key step in the core accretion theory for planet formation is the growth of small millimeter- to centimeter-sized dust grains into large meter- to kilometer-sized planetesimals \citep{Bodenheimer_Pollack_1986}. When the local dust-to-gas ratio reaches or exceeds unity (i.e. $\epsilon \geq 1.0$) the streaming instability may help dust clump together and therefore trigger planetesimal formation \citep{Youdin_Goodman_2005, Johansen_Youdin_2007}. The results of our experiment clearly show that the maximum size of dust grains capable of penetrating the gap increases with wind strength. Here, we investigate how that affects the dust-to-gas ratio throughout the disk.

\begin{figure*}
    \centering
    \includegraphics[width=0.9\linewidth]{ 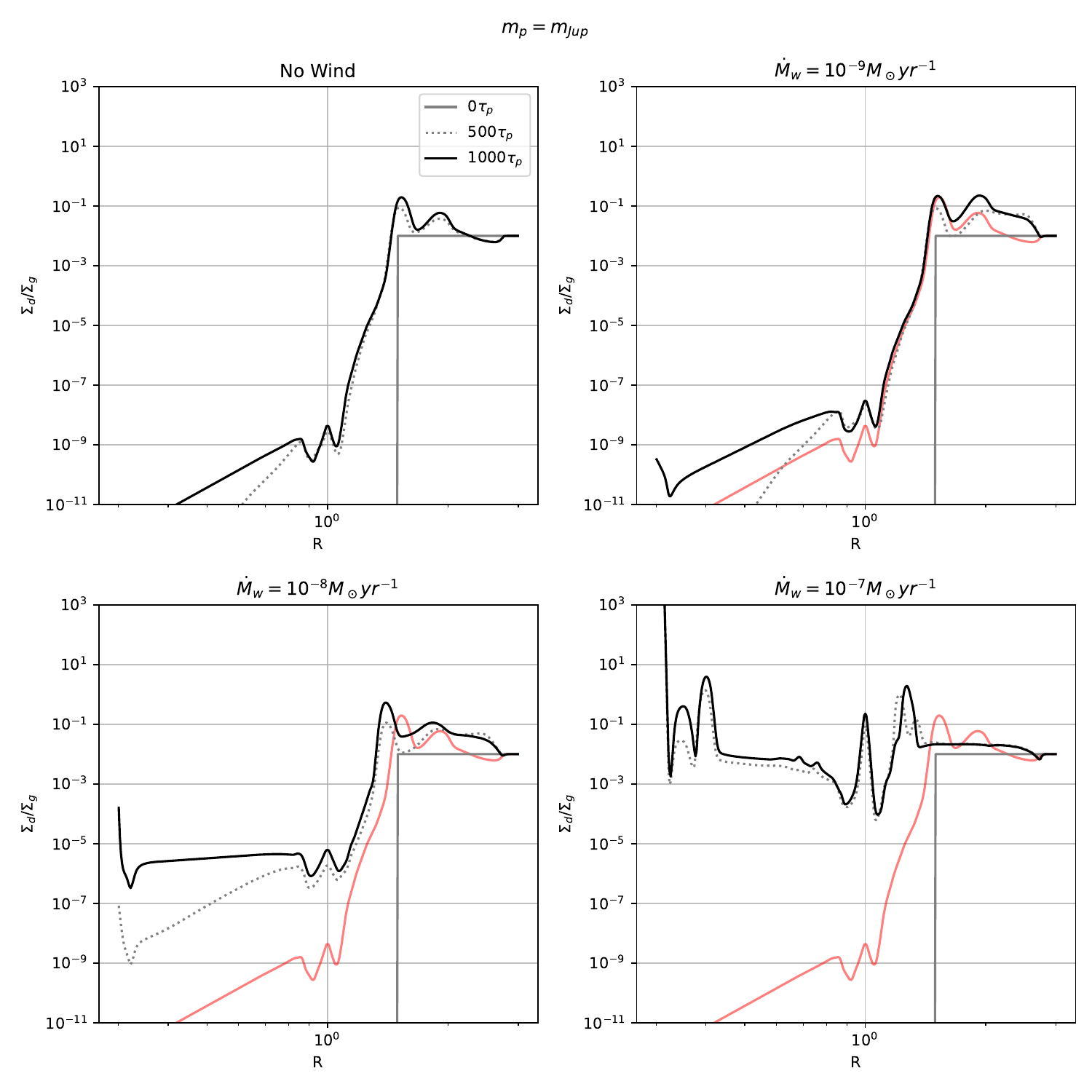}
    \caption{Evolution of the dust-to-gas ratio throughout the disk for the case of $m_p = m_{Jup}$. The dust-to-gas ratio is $\epsilon = \Sigma_d/\Sigma_g$, where $\Sigma_d$ is the total dust surface density and $\Sigma_g$ is the gas surface density. The four panels represent models with different wind mass loss rates: no wind (top left), $\dot{M}_w = 10^{-9} \rm M_\odot ~yr^{-1}$ (top right), $\dot{M}_w = 10^{-8} \rm M_\odot ~yr^{-1}$ (bottom left), and $\dot{M}_w = 10^{-7} \rm M_\odot ~yr^{-1}$ (bottom right). The final dust-to-gas ratio from the no wind case is plotted in red in each panel for a visual comparison.}
    \label{fig:dust_to_gas_ratio}
\end{figure*}

Recall, we invoke $\epsilon = 0.01$ in the outer disk at the beginning of the simulation and an MRN grain size distribution \citep[MRN;][]{MRN}. The dust-to-gas ratio throughout the disk is shown in Figure \ref{fig:dust_to_gas_ratio} for different wind strengths, this analysis was done for the case of $\rm M_p = M_{Jup}$ and we found that the result is similar for other planet masses tested (see Figure \ref{fig:dust_to_gas_ratio_all} for the planet mass parameter study). Most notably, the dust-to-gas ratio in the inner disk increases with increasing wind strength due to the enhanced influx of large dust grains which contain most of the of the dust mass when a MRN distribution is assumed. When there is no wind (Figure \ref{fig:dust_to_gas_ratio}, top left panel), $\epsilon < 10^{-9}$ in the inner disk. Then, in the case of a weak wind with $\dot{M}_w = 10^{-9} \rm M_\odot ~yr^{-1}$ (Figure \ref{fig:dust_to_gas_ratio}, top right panel), $\epsilon$ is generally higher in the inner disk then in the previous case. Next, in the intermediate wind case (Figure \ref{fig:dust_to_gas_ratio}, bottom left panel), we find the dust-to-gas ratio in the inner disk is further enhanced with $\epsilon \approx 10^{-6}$ when $\dot{M}_w = 10^{-8} \rm M_\odot ~yr^{-1}$. Finally, when the wind is strong (Figure \ref{fig:dust_to_gas_ratio}, bottom right panel), $\epsilon \approx 10^{-2}$ is even higher then the previous cases. In this case, a pressure bump develops around $R = 0.4 r_p$ where $\epsilon > 0.1$. 

To summarize, wind-driven accretion can affect the dust-to-gas ratio in the inner disk and can lead to large $\epsilon$ when the wind is strong. This indicates that wind-driven accretion could play a role in subsequent planet formation in the inner disk because the higher the dust-to-gas ratio is, the more likely planetesimals will form via the streaming instability. This has important implications for pebble isolation mass and planet formation theory since it is possible for large dust grains to be replenished in the inner disk, after the formation of a planet-opened gap, under the right windy conditions 

In the outer disk, the precise location where dust grains are trapped varies and the dust-to-gas within the trap increases with wind strength. When there is no wind, the dust trap is located around $r = 1.5 r_p$ and at this location $\epsilon = 0.22$. Then, in the case that $\dot{M}_w = 10^{-9} \rm M_\odot yr^{-1}$ the location of the dust trap is similar, but now $\epsilon = 0.24$ so it is slightly enhanced in dust compared to the previous case. In the intermediate wind case the dust trap moves inward closer to $r = 1.1 r_p$ and $\epsilon = 0.55$ within the trap. Finally, when $\dot{M}_w = 10^{-7} \rm M_\odot yr^{-1}$ the dust trap moves in even closer to the location of planet and $\epsilon = 1.64$ within the trap. In the last case, the dust-to-gas ratio within the pressure bump in the outer disk exceeds unity, indicating that this is a good place for subsequent planet formation to occur via the streaming instability (see \cite{Carrera_2021} for more on streaming instability and planetesimal formation in the pressure bump). Overall, a strong wind can push the pressure bump inward and also cause an enhancement of dust in the trap. 

\subsection{Implications for circumplanetary disk (CPD)}
\begin{figure*}
    \centering
    \includegraphics[width=1.0\linewidth]{ 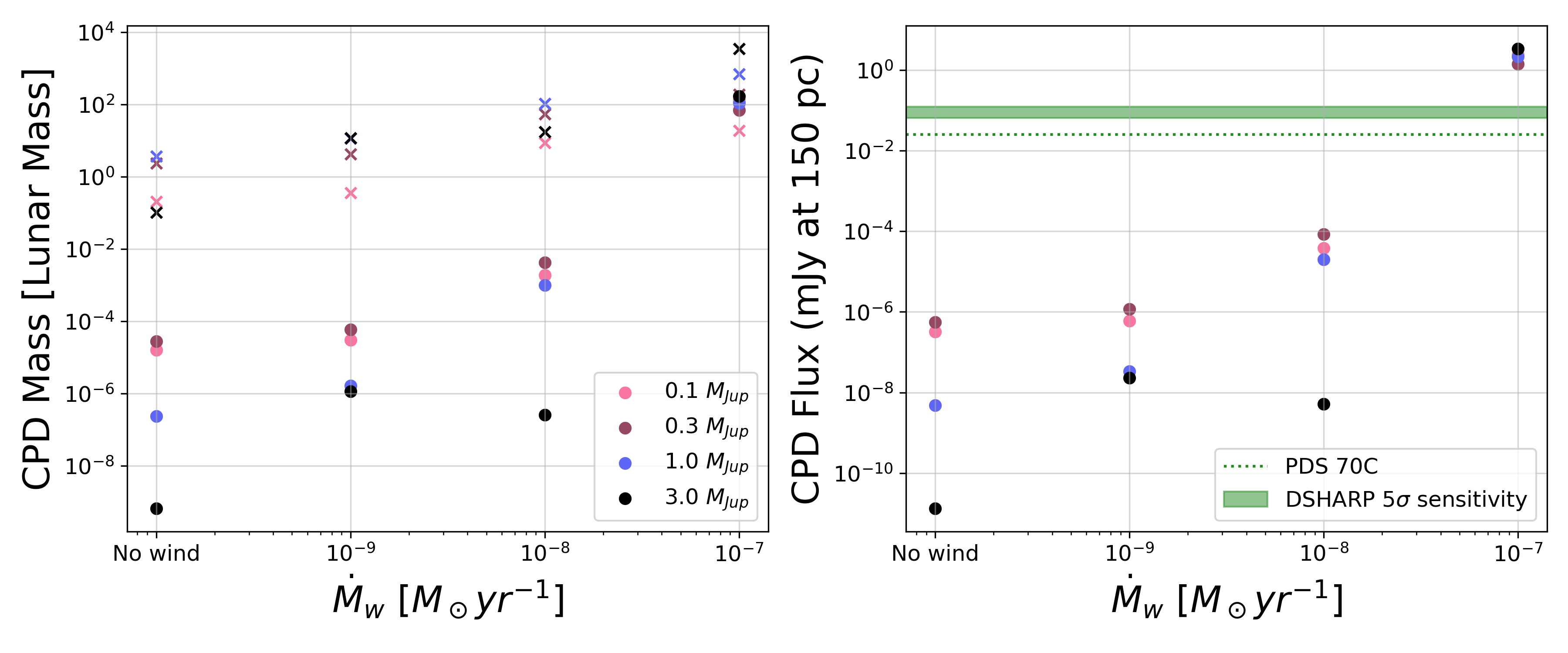}
    \caption{Left panel: Mass within 1/3 of the planets Hill radius for each of the models, calculated after 1000 orbits. The x's are gas mass and the solid points are dust mass. Right panel: Luminosity of the CPD region (corresponding to $\frac{1}{3} \rm R_{Hill}$) evaluated at $\lambda = 1.3 mm$ and assuming T=20K, $\kappa = 2.4 \rm cm^2g^{-1}$, and d=150 pc. For reference PDS 70c (rescaled to 150 pc) is shown as the dotted green line \citep{Isella_2019} and the DSHARP 5$\sigma$ detection limit \citep{Andrews_2018} is shown as the shaded green region.}
    \label{fig:CPD_massANDflux}
\end{figure*}

Pebble accretion models of planet formation predict that gas giants may become highly enriched in volatile elements as compared to refractories because giant planets induce pressure extrema in the disk that inhibit the inward migration and accretion of solids \citep{Schneider_Bitsch_A, Schneider_Bitsch_B, Schneider_Bitsch_2022}. Our results suggest that once the planet opened a deep gap, the delivery of dust onto the CPD is limited, unless the wind is strong. To investigate the effect of wind-driven accretion on the composition of the forming planet, we calculate the gas and dust mass located within $\frac{1}{3} R_{Hill}$ of the planet (see Figure \ref{fig:CPD_massANDflux} left panel), where $R_{Hill} = a \sqrt[3]{\frac{M_p}{3 (M_* + M_p)}}$ is the planets Hill radius. After a few hundred orbits, the CPD mass reaches a steady state. In most cases when the wind strength is low ($\rm \dot{M}_w \leq 10^{-8} M_{\odot}~yr^{-1}$), the dust mass in the CPD region is small, $M_{CPD} < 10^{-2}~\text{lunar masses}$, because large dust grains are trapped in the outer disk. The CPD mass tends to be larger for smaller planet masses because the smaller planets have a shallower and narrower gap which allows more dust to flow-in. When the wind is very strong ($\dot{M}_w = 10^{-7} \rm M_\odot ~yr^{-1}$) we find that the CPD dust mass is higher, reaching $M_{CPD} \gtrsim 100 ~\text{lunar masses}$ for $M_P = 0.1 M_{Jup}$. We note that the mass we report is the amount of material that is added to the CPD after the planet fully opened a gap. In reality, CPDs can have a larger amount of mass if they inherited mass before the gap is fully established.

To contextualize these results in terms of observations we also calculated the 1.3 mm flux of the CPD based on the CPD dust mass. Following the methods of \cite{miotello2022settingstageplanetformation} the flux is computed as
\begin{equation}
    F_\lambda = \frac{B_\lambda(T_d) \kappa}{d^2} M_{d},
\end{equation}
where $B_\lambda(T)$ is the Planck function, $\kappa$ is the dust opacity, $d$ is the distance to the source, and $M_{d}$ is the CPD dust mass. To be consistent with literature estimates, such as \cite{Andrews_2021}, we evaluate the flux at $\lambda = 1.3 mm$ and assume T=20K, $\kappa = 2.4 \rm cm^2g^{-1}$, and d=150 pc. The right panel of Figure \ref{fig:CPD_massANDflux} shows the CPD flux for each model in this study. In this figure we also include the flux of PDS 70c's CPD \citep{Isella_2019} and the $5\sigma$ sensitivity of the DSHARP survey \citep{Andrews_2018}. When there is no wind, or the wind is weak, there is a small amount of dust in the CPD so it is not very bright with $F < 10^{-4} \rm mJy$, which may explain a large number of non-detection from previous mm continuum observations \citep[e.g.,][]{Andrews_2021,curone2025}. The CPD is brighter in disks with higher wind strength due to the increased dust mass; for the smallest planet mass, $F = 2.40 \rm mJy$ which is above the flux of PDS 70c and the upper limit of the DSHARP survey \citep{Andrews_2021}.

\subsection{Implications for Solar System formation and inner disk chemistry}\label{subsec:Implications}
There is meteoritic evidence that Jupiter formed early in the Solar System, within 1 million years, and separated the solar nebula into two parts \citep{Kruijer_2017}. Commonly referred to as the NC-CC dichotomy, this evidence suggests that the early formation of Jupiter sufficiently blocked pebble flow into the inner Solar System \citep{Kruijer_2020, Kleine_2020}. Our results suggest that the solar nebula should not have had strong winds that could enable large, chondrule-sized grains to penetrate the gap opened by Jupiter and flow into the inner solar nebula. 

On the other hand, recent observations of protoplanetary disks with James Webb Space Telescope have revealed that compact disks without known substructures have excess water emission compared to large disks with substructures \citep{Banzatti_2023, Temmink_2024}. It has been proposed that this may be evidence of more efficient pebble drift in disks without substructures since water needs to be delivered via pebbles; once pebbles pass the snowline, water will be released in vapor form that is observable \citep{kalyaan2023effectdustevolutiontraps, Houge_2025}. However, the picture is somewhat more complicated as there are disks that have deep gaps and are also rich in warm water in the inner disk, which indicates that substructures may not be fully blocking the pebble flow (e.g., \citealt{Gasman_2025}). One possible explanation for this, in the context of our study, is that winds in those disks may have enabled large grains to overcome pressure bumps and drift into the inner disk. This could potentially help explain why warm-H$_2$O is observed in some protoplanetary disks with substructures. To prove or dispute this idea, it will be interesting to see if there is any positive correlation between the amount of warm water in the inner disk and MHD-wind outflow rate.

\subsection{Caveats}\label{subsec:caveats}
Although our simulations demonstrated the possibility that magnetically-driven winds can enhance pebble drift and potentially allow pebbles to penetrate planet-opened gaps, future simulations are warranted to further explore this possibility. To this end, in this section we discuss limitations to our model and future directions.

First of all, we implemented a parameterized wind model in a two-dimensional computational domain. In three-dimensional non-ideal MHD simulations, wind-driven accretion flows generally form at $1-2$ scale heights from the disk midplane \citep[e.g.,][]{Wafflard_Lesur_2023, Wafflard_Lesur_2025,hu_2024}. This suggests that, in order for dust grains to penetrate a planet-induced gap, they need to be lofted to where wind-driven accretion flows exist. Three-dimensional hydrodynamic planet-disk interaction simulations showed that it is possible to have large, mm-sized grains lofted to a gas scale height through the turbulence and meridional flows driven by the spiral wave instability \citep{bae2016b,bae2016c,bi2021,binkert2021}. Yet, whether large, mm-sized grains can be indeed lofted high enough to take advantage of rapid inward gas flows needs to be tested with three-dimensional non-ideal MHD simulations. Related, our parameterized wind model imposes radial velocities that are azimuthally symmetric and constant over time (Equation \ref{eqn:u_r_wind}). In reality, wind-driven flows must be azimuthal structures and time variable. Again, interaction between dust grains and more realistic wind-driven accretion flows need to be studied using three-dimensional non-ideal MHD simulations.

Our simulations do not include dust size evolution. Including dust coagulation and fragmentation can change the maximum grain size as well as the exact size distribution \citep{Drazkowska_2019}, which determine how much solid mass can penetrate planet-induced gaps. While we defer including dust size evolution to a future study, we estimate the maximum grain size in our initial disk. Using the formula from \citep{birnstiel2012} 
\begin{equation}
    a_{\rm frag} = {2 \over 3\pi} {\Sigma_g \over \rho_{\rm int} \alpha} {v_{\rm frag}^2 \over c_s^2}
\end{equation}
and adopting $\Sigma_g = 2.11~{\rm g~cm}^{-2}$, $\rho_{\rm int} = 3~{\rm g~cm}^{-3}$, $\alpha=10^{-3}$, $(h/r)_p = 0.05$ which results in $c_s = 272~{\rm m~s}^{-1}$, and the fragmentation velocity of grains $v_{\rm frag} = 10~{\rm m~s}^{-1}$, the fragmentation-dominated maximum grain size at the location of the planet in our initial disk is about 2~mm. 

Radial drift is known to be another limiting factor in determining the maximum grain size. However, at the time we insert grains, planets already formed a strong pressure bump beyond their orbits so in the absence of strong winds, radial drift should not be a limiting factor as we showed in Figure \ref{fig:b_crit}.
\begin{enumerate}
    \item In conclusion, given that our largest grain size is comparable to the fragmentation-dominated maximum grain size, we believe that including dust size evolution would not change our results drastically.
\end{enumerate}

Lastly, it will be interesting to study how much dust wind-driven accretion can provide to the inner disk when there is more than one giant planet in the disk. This will be particularly interesting in the context of solar system formation as Jupiter and Saturn could have opened a common gap in the solar nebula \citep[e.g.,][]{walsh2011}. 

\section{Conclusion}\label{sec:conclusion}
In this study, we conducted a numerical simulation experiment to investigate how multifluid material flows from the outer disk to the inner disk in a 2D protoplanetary disk. For this purpose, we implemented a parameterized version of the magnetic-wind model and compared the results with pure viscous accretion. Some of the main results of the experiment are summarized here:

\begin{itemize}
    \item When the disk evolves under pure viscous accretion, we found that small grains can penetrate the planet-induced gap and flow into the inner disk, whereas large grains are filtered by the planet-induced gap and trapped in the outer disk. The size of grains that are capable of penetrating the gap decreases with planet mass. This is consistent with previous studies \citep[e.g.,][]{Zhu_2012,Weber_2018,bae2019,huang2025leakydusttrapsplanetembedded}. 
    \item When the disk evolves with a magnetically-driven wind, the planet-induced gap becomes more permeable and the maximum size of dust grains capable of penetrating the gap increases with the wind strength. This result holds for all planet masses tested.
    \item A strong wind (i.e. $\dot{M}_w = 10^{-7} \rm M_\odot yr^{-1}$) can cause large mm-sized dust grains to overcome the dust trap in the outer disk and flow into the inner disk. 
    \item We computed a depletion factor which compares the inner disk dust mass of each model to a reference model with no planet and no wind. When there is no wind, large mm-sized grains are more depleted in the inner disk than small $\mu$m-sized grains. Then when there is a magnetically-driven wind, dust grains are generally less depleted in the inner disk than in the no wind case. When the wind is strong ($\dot{M}_w = 10^{-7} \rm M_\odot yr^{-1}$), mm-sized grains are enhanced in the inner disk compared to the reference model where no wind is present.
    \item Disk winds affect how material flows in the disk, leading to asymmetric flows in the co-rotating region. The flow patterns are broadly consistent with non-ideal MHD simulations \citep[e.g.,,][]{Wafflard_Lesur_2023, Wafflard_Lesur_2025, hu_2024}.
\end{itemize}

Overall, we find that magnetically-driven winds may enhance pebble drift through the planet-opened gap into the inner disk when the wind mass loss rate is sufficiently large. Future studies incorporating self-consistently-driven magnetized winds in three dimensions will prove (or dispute) this possibility.

\begin{acknowledgments}
This material is based upon work supported by the National Science Foundation Graduate Research Fellowship under Grant No. DGE-2236414. Any opinion, findings, and conclusions or recommendations expressed in this material are those of the authors(s) and do not necessarily reflect the views of the National Science Foundation. The authors thank Xiao Hu for helpful discussion. The authors acknowledge University of Florida Research Computing for providing computational resources and support that have contributed to the research results reported in this publication (\url{http://www.rc.ufl.edu}.)
\end{acknowledgments}

%

\vspace{5mm}


\software{astropy \citep{2013A&A...558A..33A,2018AJ....156..123A},
          matplotlib \citep{Hunter:2007},
          numpy \citep{harris2020array},
          pandas \citep{reback2020pandas},
          scipy \citep{2020SciPy-NMeth},
          FARGO3D \citep{fargo_multifluid, fargo3d}
          }



\appendix

\section{Smoothing Length Test} \label{sec:smoothing_length}
\begin{figure*}
    \centering
    \includegraphics[width=\textwidth]{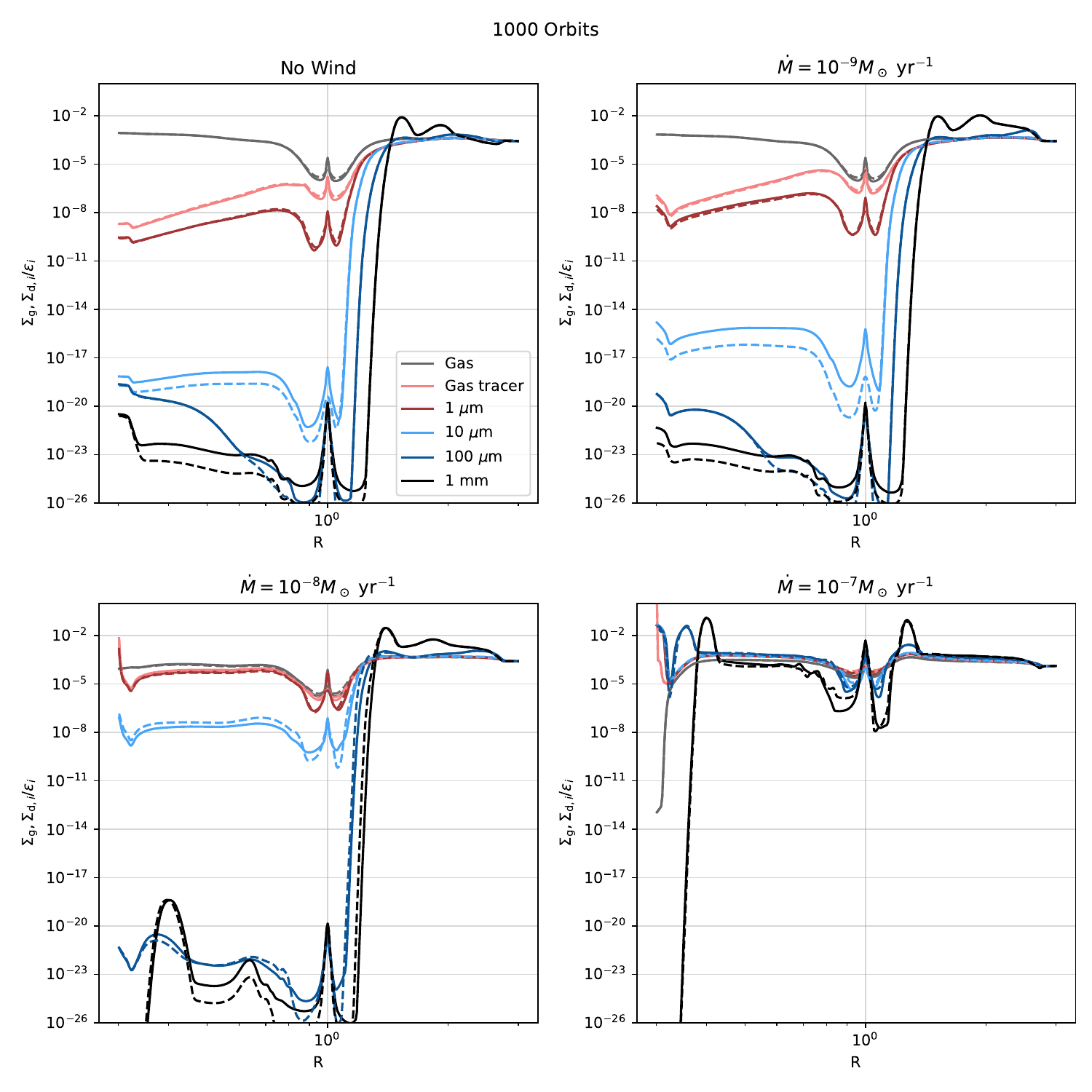}
    \caption{Azimuthally-averaged surface density profiles after 1000 orbits of evolution for $M_p = M_{Jup}$ models with different wind strengths. The solid curves are from simulations with $s=0.22h_p$ while the dashed curved are from simulations with $s=0.6h_p$. Each gas and dust fluid is plotted corresponding to the colors in the figure legend. The surface density has been scaled up from the true surface density by a factor of $\epsilon_i$.}
    \label{fig:smoothingLength}
\end{figure*}

For our fiducial models, we adopted $s = 0.22h_p$ to avoid over-smoothing the planet's gravity. Yet, to ensure that our results are not sensitive to the choice of smoothing lengths, we ran additional simulations with $s = 0.6 h_p$ for the model with $m_p = 1 m_{Jup}$. Figure \ref{fig:smoothingLength} shows the results of these additional simulations, along with the fiducial model where $s = 0.22h_p$ is used. As shown in the figure, the overall radial profiles of the gas and dust surface densities are not sensitive to the smoothing length. We found that there is no significant difference in the dust density in the inner disk when winds are strong ($\dot{M} = 10^{-7}~M_\odot~{\rm yr}^{-1}$). When winds are weak or absent, we found that  the dust density in the inner disk is generally higher with a smaller smoothing length of $s = 0.22h_p$, by a factor of up to $\sim10$. However, since the inner disk is highly devoid of large grains when winds are weak, this difference does not affect our overall conclusions.

\section{Supplementary Figures}

To avoid redundancy, we present the results of the Jupiter mass planet case in the main body of the paper and provide the other planet masses here. Figures \ref{fig:surfaceDensImage_0P1Jup}, \ref{fig:surfaceDensImage_Sat}, and \ref{fig:surfaceDensImage_3Jup} show the resulting 2D surface density distribution after 1000 orbits under varying levels of wind strength for the $M_P = 0.1 M_{Jup}$, $0.3 M_{Jup}$, and $3 M_{Jup}$ respectively. The main results are summarized in Section \ref{subsec:planet_mass_param_study}.

In Section \ref{subsec:morphology_massFlux} we present the total mass flux at $\rm R=0.9 r_p$ for the case of $\rm M_p = 1.0~M_{Jup}$ and show that the mass flux is always higher when $\dot{M}_w = 10^{-7} M_\odot~{\rm yr}^{-1}$ then the case with no wind. Figure \ref{fig:massFlux_windStrength_grainSize_all} shows the same result but for the planet mass parameter study. For the gas we see a similar trend for every planet mass tested; the gas flows in at a higher rate for larger $\dot{M}_w$. For the small dust, when there is no wind, generally the mass flux is lower for higher mass planets. This is because the smaller mass planets form "leakier" gaps due to the gap being shallower and narrower then the more massive planets (see Figure \ref{fig:SD_allMp}). In other words, massive planets are more efficient at trapping dust in the outer disk then the less massive planets. In all cases, the mass flux of the largest dust grains is highest when $\dot{M}_w = 10^{-7} M_\odot~{\rm yr}^{-1}$ and for the other wind strengths it is negligible. Indicating that for every planet mass tested, large mm-sized grains are trapped in the outer disk except when $\dot{M}_w = 10^{-7} M_\odot~{\rm yr}^{-1}$. For every planet mass tested the overall result remains that--after the formation of a planet-opened gap--a strong enough wind can increase the gas mass flux and therefore push material into the inner disk that would otherwise be trapped in the outer disk under pure viscous accretion.

In Section \ref{subsec:dtg_ratio} we present the evolution of the azimuthally averaged dust-to-gas ratio ($\epsilon = \Sigma_d/\Sigma_g$) throughout the disk for the case of $\rm M_p = 1.0~M_{Jup}$. Figure \ref{fig:dust_to_gas_ratio_all} shows the same result for the planet mass parameter study. One distinction is that the dust-to-gas ratio in the inner disk is slightly higher for the $\rm M_p = 0.1~M_{Jup}$ case ($\epsilon = 0.004$) because the gap is leakier than the more massive planets. Qualitatively the result is the same for every planet mass tested: in the inner disk and gap region $\epsilon$ increases with increasing $\dot{M}_w$, in the outer disk the location of the dust trap moves inward for increasing $\dot{M}_w$, and the trap is enhanced in dust for larger $\dot{M}_w$.

\begin{figure*}
    \centering
    \includegraphics[width=\textwidth]{ 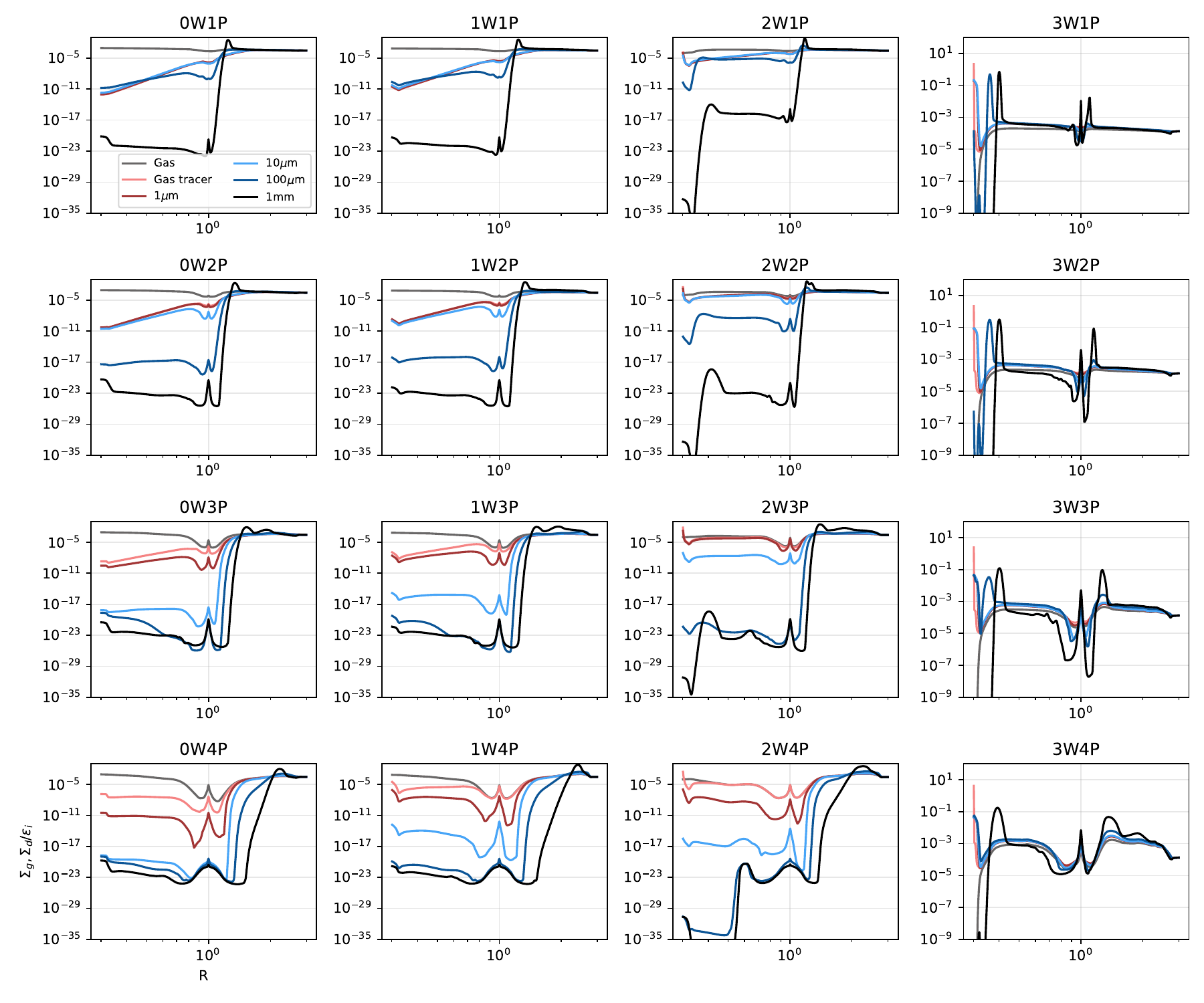}
    \caption{Surface density radial profile after 1000 orbits of evolution for every model in this study. From top to bottom the panels show the result for $\rm M_p = 0.1~M_{Jup}, 0.3~M_{Jup}, 1.0~M_{Jup}$, and $\rm 3.0~M_{Jup}$, respectively. From right to left the panels show the result for different wind mass loss rates: no wind, $\dot{M}_w = 10^{-9} \rm M_\odot yr^{-1}$, $\dot{M}_w = 10^{-8} \rm M_\odot yr^{-1}$, and $\dot{M}_w = 10^{-7} \rm M_\odot yr^{-1}$. The dust surface density is scaled up from the true density according to $\epsilon_i$.}
    \label{fig:surfaceDensProfile_all}
\end{figure*}

\begin{figure*}
    \centering
    \includegraphics[width=\textwidth]{ 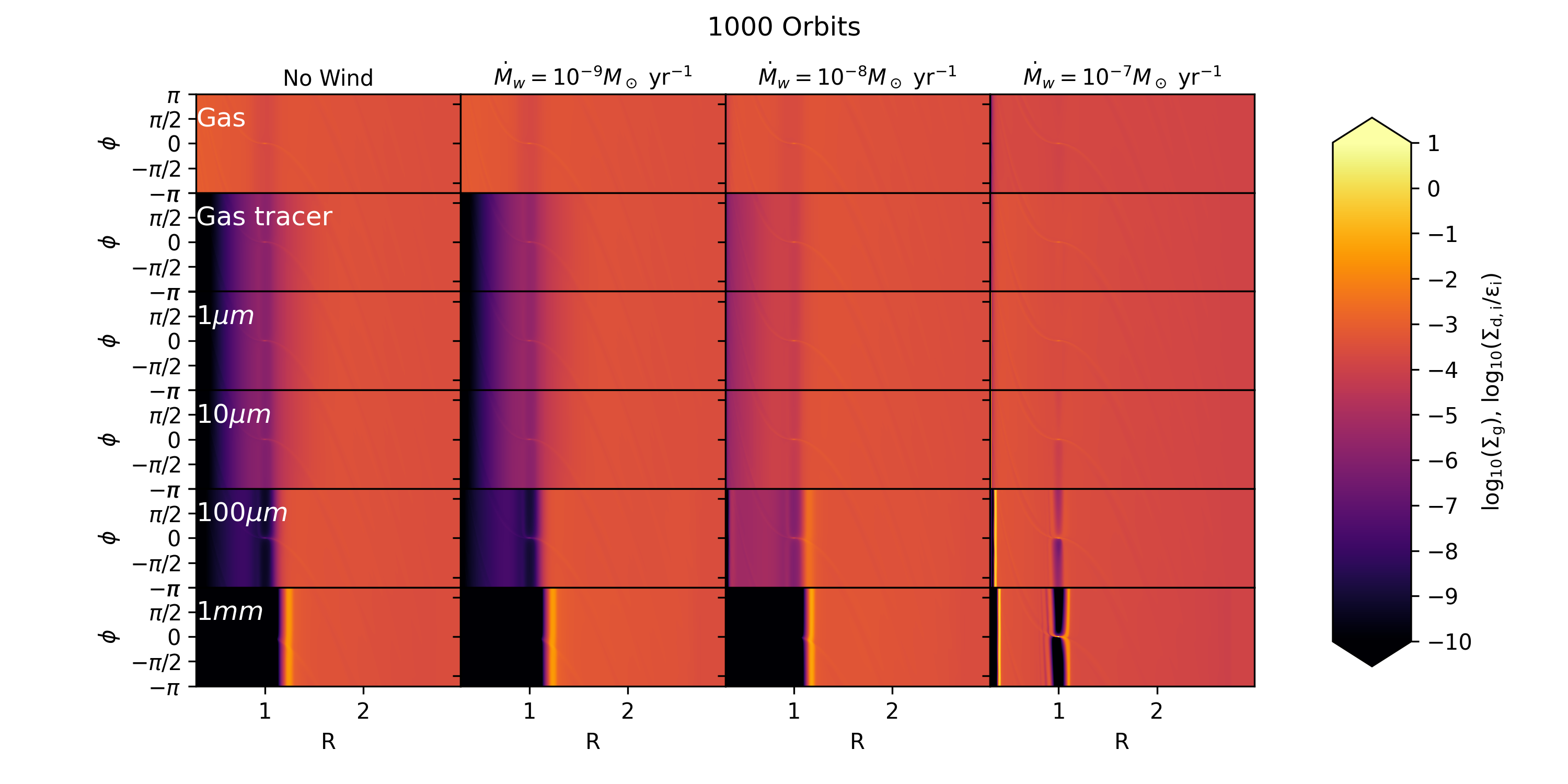}
    \caption{Surface density profile after 1000 orbits in polar coordinates for the m$_P$ = 0.1 x Jupiter mass case. The planet is located at $(r, \phi) = (1.0, 0.0)$. The rows are the result of different wind mass loss parameters, increasing from left to right: no wind, $b \approx 10^{-3}$, $b \approx 10^{-4}$, and $b \approx 10^{-5}$. The columns are different fluids in the disk, increasing from top to bottom: gas, gas tracer (St = $1 \times 10^{-6}$), 1 $\mu$m, 10 $\mu$m, 100 $\mu$m, and 1 mm. The dust surface density is scaled up from the true density according to $\epsilon_i$.}
    \label{fig:surfaceDensImage_0P1Jup}
\end{figure*}

\begin{figure*}
    \centering
    \includegraphics[width=\textwidth]{ 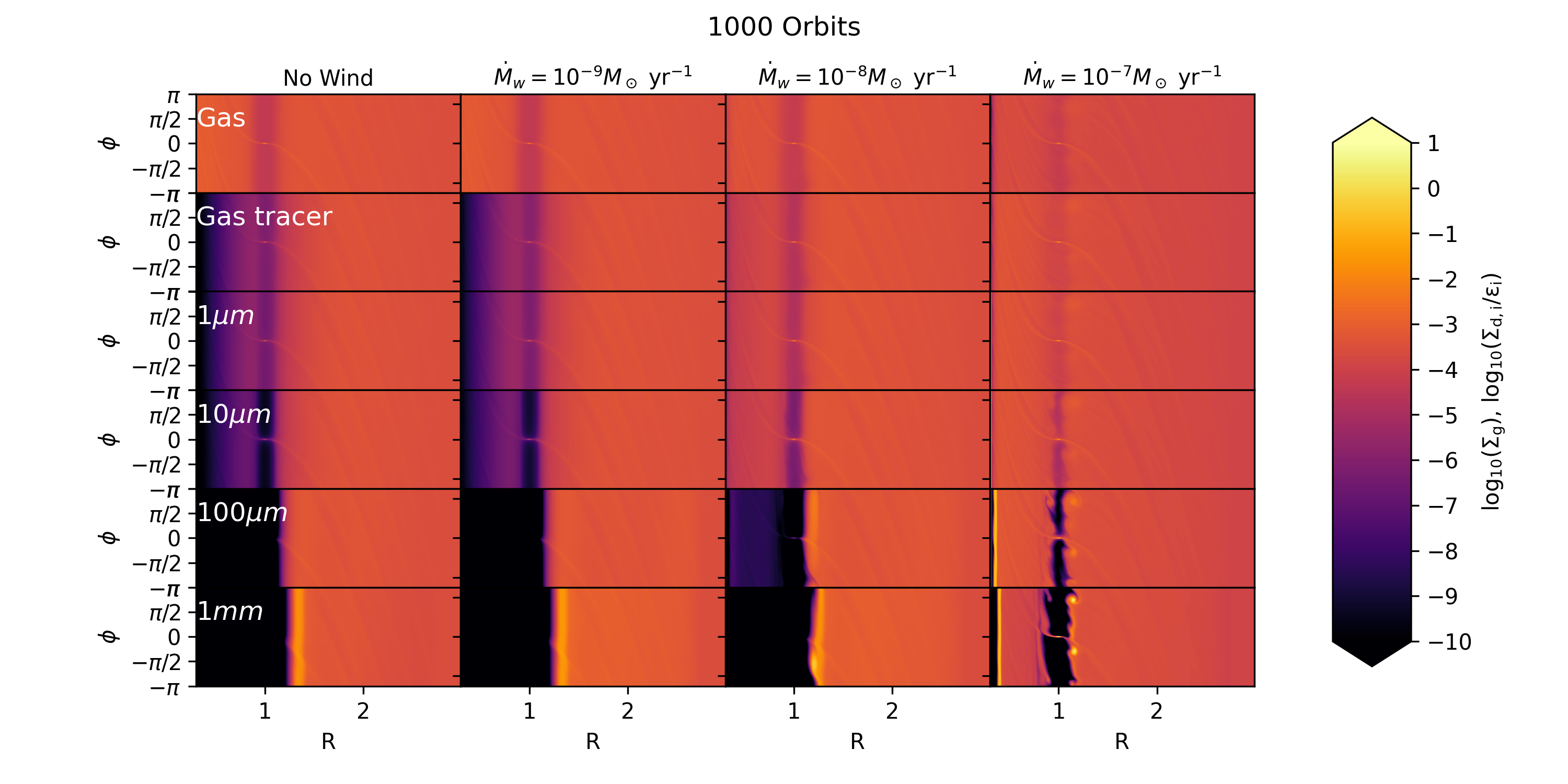}
    \caption{Surface density profile after 1000 orbits in polar coordinates for the m$_P$ = Saturn mass case. The planet is located at $(r, \phi) = (1.0, 0.0)$. The rows are the result of different wind mass loss parameters, increasing from left to right: no wind, $b \approx 10^{-3}$, $b \approx 10^{-4}$, and $b \approx 10^{-5}$. The columns are different fluids in the disk, increasing from top to bottom: gas, gas tracer (St = $1 \times 10^{-6}$), 1 $\mu$m, 10 $\mu$m, 100 $\mu$m, and 1 mm. The dust surface density is scaled up from the true density according to $\epsilon_i$.}
    \label{fig:surfaceDensImage_Sat}
\end{figure*}

\begin{figure*}
    \centering
    \includegraphics[width=\textwidth]{ 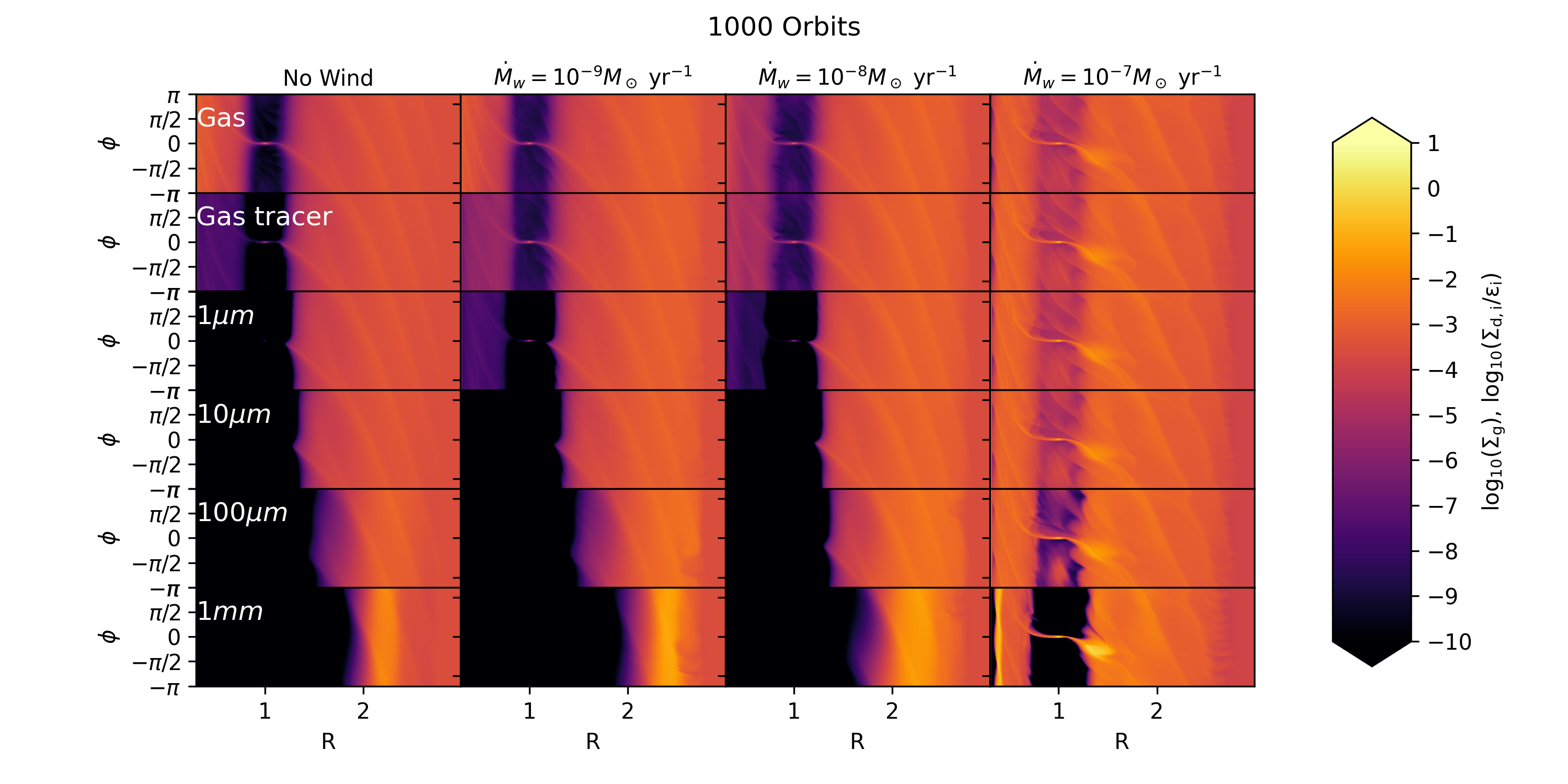}
    \caption{Surface density profile after 1000 orbits in polar coordinates for the m$_P$ = 3 x Jupiter mass case. The planet is located at $(r, \phi) = (1.0, 0.0)$. The rows are the result of different wind mass loss parameters, increasing from left to right: no wind, $b \approx 10^{-3}$, $b \approx 10^{-4}$, and $b \approx 10^{-5}$. The columns are different fluids in the disk, increasing from top to bottom: gas, gas tracer (St = $1 \times 10^{-6}$), 1 $\mu$m, 10 $\mu$m, 100 $\mu$m, and 1 mm. The dust surface density is scaled up from the true density according to $\epsilon_i$.}
    \label{fig:surfaceDensImage_3Jup}
\end{figure*}

\begin{figure*}
    \centering
    \includegraphics[width=1.0\linewidth]{ 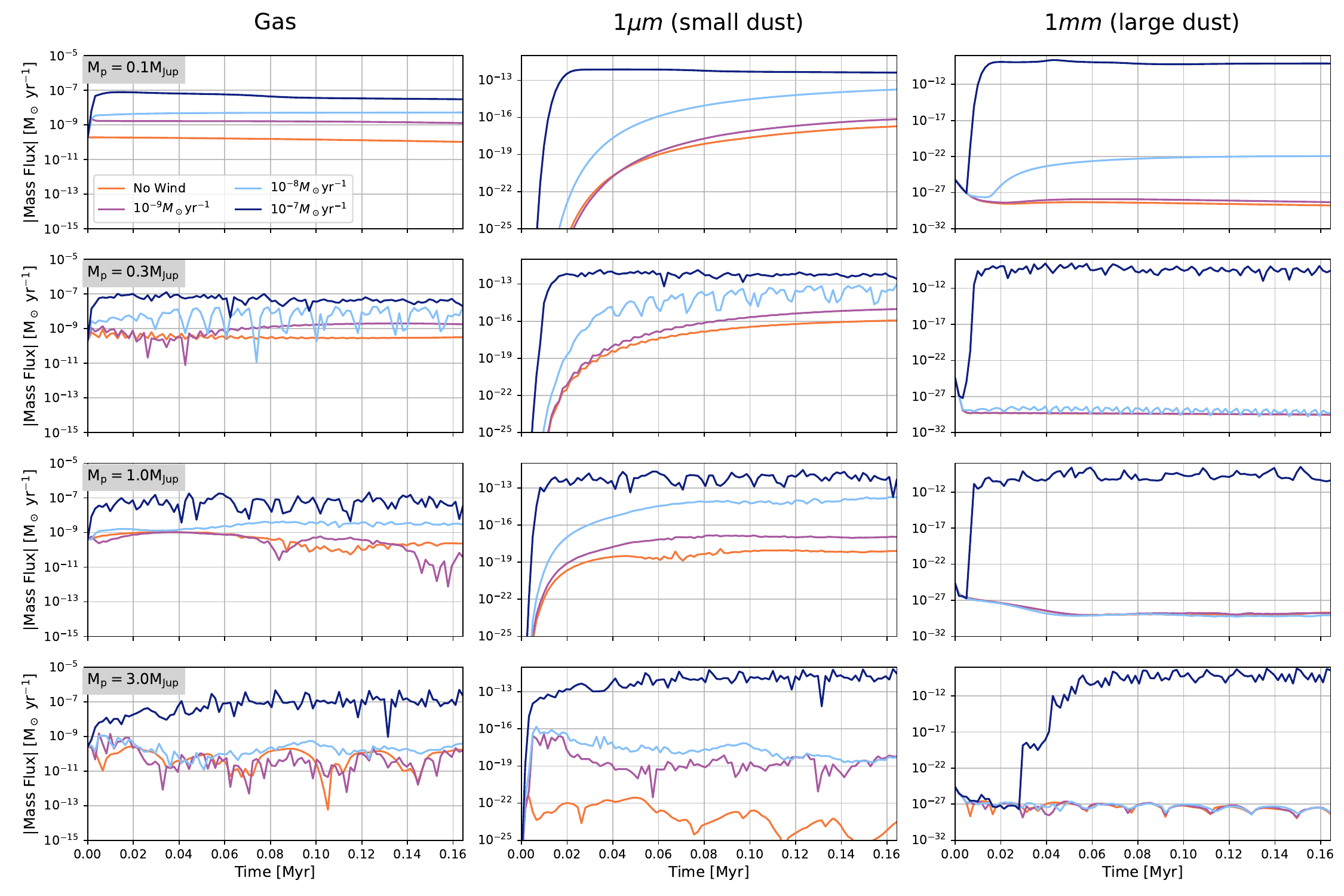}
    \caption{Absolute value of the total mass flux at $\rm R=0.9 r_p$ over time for all the planet masses in this study. The different colors show different wind strengths corresponding to the figure legend. From top to bottom the panels show the result for $\rm M_p = 0.1~M_{Jup}, 0.3~M_{Jup}, 1.0~M_{Jup}$, and $\rm 3.0~M_{Jup}$, respectively. From right to left the panels show the result for gas, $1 \mu m$ dust, and $1 mm$ dust. The mass flux toward in inner disk is always higher when $\dot{M}_w = 10^{-7} \rm M_\odot yr^{-1}$ compared to the case with no wind.}
    \label{fig:massFlux_windStrength_grainSize_all}
\end{figure*}

\begin{figure*}
    \centering
    \includegraphics[width=1.0\linewidth]{ 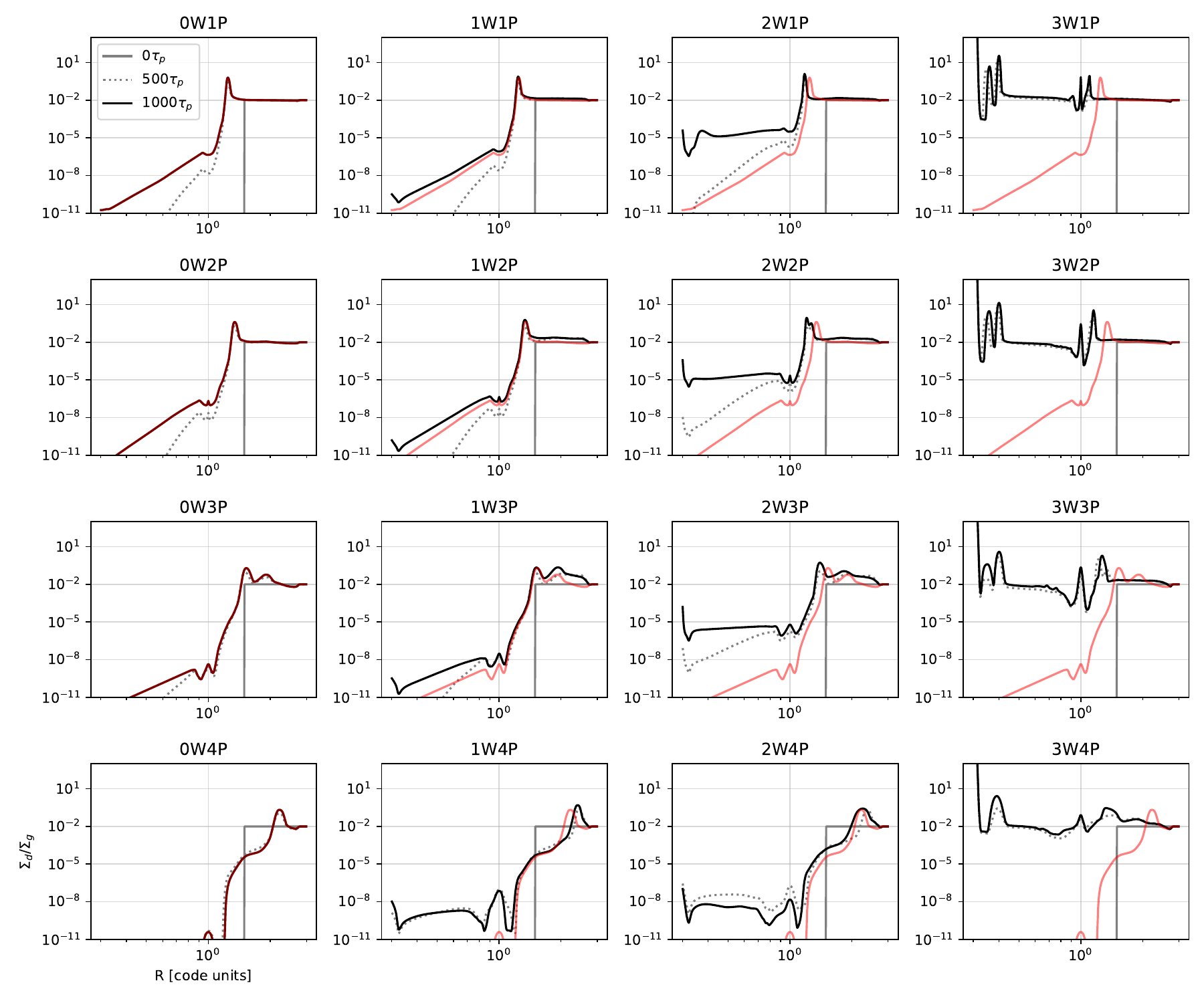}
    \caption{Evolution of the azimuthally averaged dust-to-gas ratio throughout the disk for all the planet masses in this study. The dust-to-gas ratio is $\epsilon = \Sigma_d/\Sigma_g$, where $\Sigma_d$ is the total dust surface density and $\Sigma_g$ is the gas surface density. From top to bottom the panels show the result for $\rm M_p = 0.1~M_{Jup}, 0.3~M_{Jup}, 1.0~M_{Jup}$, and $\rm 3.0~M_{Jup}$, respectively. From right to left the panels show the result for different wind mass loss rates: no wind, $\dot{M}_w = 10^{-9} \rm M_\odot yr^{-1}$, $\dot{M}_w = 10^{-8} \rm M_\odot yr^{-1}$, and $\dot{M}_w = 10^{-7} \rm M_\odot yr^{-1}$. The final dust-to-gas ratio from the no wind case is plotted in red in each panel for a visual comparison.}
    \label{fig:dust_to_gas_ratio_all}
\end{figure*}


\bibliography{sample631}{}

\begin{thebibliography}{}
\expandafter\ifx\csname natexlab\endcsname\relax\def\natexlab#1{#1}\fi
\providecommand{\url}[1]{\href{#1}{#1}}
\providecommand{\dodoi}[1]{doi:~\href{http://doi.org/#1}{\nolinkurl{#1}}}
\providecommand{\doeprint}[1]{\href{http://ascl.net/#1}{\nolinkurl{http://ascl.net/#1}}}
\providecommand{\doarXiv}[1]{\href{https://arxiv.org/abs/#1}{\nolinkurl{https://arxiv.org/abs/#1}}}

\bibitem[{{Andrews}(2020)}]{andrews2020}
{Andrews}, S.~M. 2020, \araa, 58, 483, \dodoi{10.1146/annurev-astro-031220-010302}

\bibitem[{{Andrews} {et~al.}(2018){Andrews}, {Huang}, {P{\'e}rez}, {Isella}, {Dullemond}, {Kurtovic}, {Guzm{\'a}n}, {Carpenter}, {Wilner}, {Zhang}, {Zhu}, {Birnstiel}, {Bai}, {Benisty}, {Hughes}, {{\"O}berg}, \& {Ricci}}]{Andrews_2018}
{Andrews}, S.~M., {Huang}, J., {P{\'e}rez}, L.~M., {et~al.} 2018, \apjl, 869, L41, \dodoi{10.3847/2041-8213/aaf741}

\bibitem[{Andrews {et~al.}(2021)Andrews, Elder, Zhang, Huang, Benisty, Kurtovic, Wilner, Zhu, Carpenter, Pérez, Teague, Isella, \& Ricci}]{Andrews_2021}
Andrews, S.~M., Elder, W., Zhang, S., {et~al.} 2021, The Astrophysical Journal, 916, 51, \dodoi{10.3847/1538-4357/ac00b9}

\bibitem[{{Astropy Collaboration} {et~al.}(2013){Astropy Collaboration}, {Robitaille}, {Tollerud}, {Greenfield}, {Droettboom}, {Bray}, {Aldcroft}, {Davis}, {Ginsburg}, {Price-Whelan}, {Kerzendorf}, {Conley}, {Crighton}, {Barbary}, {Muna}, {Ferguson}, {Grollier}, {Parikh}, {Nair}, {Unther}, {Deil}, {Woillez}, {Conseil}, {Kramer}, {Turner}, {Singer}, {Fox}, {Weaver}, {Zabalza}, {Edwards}, {Azalee Bostroem}, {Burke}, {Casey}, {Crawford}, {Dencheva}, {Ely}, {Jenness}, {Labrie}, {Lim}, {Pierfederici}, {Pontzen}, {Ptak}, {Refsdal}, {Servillat}, \& {Streicher}}]{2013A&A...558A..33A}
{Astropy Collaboration}, {Robitaille}, T.~P., {Tollerud}, E.~J., {et~al.} 2013, \aap, 558, A33, \dodoi{10.1051/0004-6361/201322068}

\bibitem[{{Astropy Collaboration} {et~al.}(2018){Astropy Collaboration}, {Price-Whelan}, {Sip{\H{o}}cz}, {G{\"u}nther}, {Lim}, {Crawford}, {Conseil}, {Shupe}, {Craig}, {Dencheva}, {Ginsburg}, {VanderPlas}, {Bradley}, {P{\'e}rez-Su{\'a}rez}, {de Val-Borro}, {Aldcroft}, {Cruz}, {Robitaille}, {Tollerud}, {Ardelean}, {Babej}, {Bach}, {Bachetti}, {Bakanov}, {Bamford}, {Barentsen}, {Barmby}, {Baumbach}, {Berry}, {Biscani}, {Boquien}, {Bostroem}, {Bouma}, {Brammer}, {Bray}, {Breytenbach}, {Buddelmeijer}, {Burke}, {Calderone}, {Cano Rodr{\'\i}guez}, {Cara}, {Cardoso}, {Cheedella}, {Copin}, {Corrales}, {Crichton}, {D'Avella}, {Deil}, {Depagne}, {Dietrich}, {Donath}, {Droettboom}, {Earl}, {Erben}, {Fabbro}, {Ferreira}, {Finethy}, {Fox}, {Garrison}, {Gibbons}, {Goldstein}, {Gommers}, {Greco}, {Greenfield}, {Groener}, {Grollier}, {Hagen}, {Hirst}, {Homeier}, {Horton}, {Hosseinzadeh}, {Hu}, {Hunkeler}, {Ivezi{\'c}}, {Jain}, {Jenness}, {Kanarek}, {Kendrew}, {Kern}, {Kerzendorf}, {Khvalko}, {King}, {Kirkby}, {Kulkarni},
  {Kumar}, {Lee}, {Lenz}, {Littlefair}, {Ma}, {Macleod}, {Mastropietro}, {McCully}, {Montagnac}, {Morris}, {Mueller}, {Mumford}, {Muna}, {Murphy}, {Nelson}, {Nguyen}, {Ninan}, {N{\"o}the}, {Ogaz}, {Oh}, {Parejko}, {Parley}, {Pascual}, {Patil}, {Patil}, {Plunkett}, {Prochaska}, {Rastogi}, {Reddy Janga}, {Sabater}, {Sakurikar}, {Seifert}, {Sherbert}, {Sherwood-Taylor}, {Shih}, {Sick}, {Silbiger}, {Singanamalla}, {Singer}, {Sladen}, {Sooley}, {Sornarajah}, {Streicher}, {Teuben}, {Thomas}, {Tremblay}, {Turner}, {Terr{\'o}n}, {van Kerkwijk}, {de la Vega}, {Watkins}, {Weaver}, {Whitmore}, {Woillez}, {Zabalza}, \& {Astropy Contributors}}]{2018AJ....156..123A}
{Astropy Collaboration}, {Price-Whelan}, A.~M., {Sip{\H{o}}cz}, B.~M., {et~al.} 2018, \aj, 156, 123, \dodoi{10.3847/1538-3881/aabc4f}

\bibitem[{{Bae} {et~al.}(2023){Bae}, {Isella}, {Zhu}, {Martin}, {Okuzumi}, \& {Suriano}}]{bae2023}
{Bae}, J., {Isella}, A., {Zhu}, Z., {et~al.} 2023, in Astronomical Society of the Pacific Conference Series, Vol. 534, Protostars and Planets VII, ed. S.~{Inutsuka}, Y.~{Aikawa}, T.~{Muto}, K.~{Tomida}, \& M.~{Tamura}, 423, \dodoi{10.48550/arXiv.2210.13314}

\bibitem[{{Bae} {et~al.}(2016{\natexlab{a}}){Bae}, {Nelson}, \& {Hartmann}}]{bae2016c}
{Bae}, J., {Nelson}, R.~P., \& {Hartmann}, L. 2016{\natexlab{a}}, \apj, 833, 126, \dodoi{10.3847/1538-4357/833/2/126}

\bibitem[{{Bae} {et~al.}(2016{\natexlab{b}}){Bae}, {Nelson}, {Hartmann}, \& {Richard}}]{bae2016b}
{Bae}, J., {Nelson}, R.~P., {Hartmann}, L., \& {Richard}, S. 2016{\natexlab{b}}, \apj, 829, 13, \dodoi{10.3847/0004-637X/829/1/13}

\bibitem[{{Bae} {et~al.}(2019){Bae}, {Zhu}, {Baruteau}, {Benisty}, {Dullemond}, {Facchini}, {Isella}, {Keppler}, {P{\'e}rez}, \& {Teague}}]{bae2019}
{Bae}, J., {Zhu}, Z., {Baruteau}, C., {et~al.} 2019, \apjl, 884, L41, \dodoi{10.3847/2041-8213/ab46b0}

\bibitem[{{Bai}(2017)}]{Bai_2017}
{Bai}, X.-N. 2017, \apj, 845, 75, \dodoi{10.3847/1538-4357/aa7dda}

\bibitem[{{Bai} \& {Stone}(2013)}]{Bai_Stone_2013}
{Bai}, X.-N., \& {Stone}, J.~M. 2013, \apj, 769, 76, \dodoi{10.1088/0004-637X/769/1/76}

\bibitem[{{Balbus} \& {Hawley}(1991)}]{Balbus_Hawley_1991}
{Balbus}, S.~A., \& {Hawley}, J.~F. 1991, \apj, 376, 214, \dodoi{10.1086/170270}

\bibitem[{{Banzatti} {et~al.}(2023){Banzatti}, {Pontoppidan}, {Carr}, {Jellison}, {Pascucci}, {Najita}, {Romero-Mirza}, {{\"O}berg}, {Kalyaan}, {Pinilla}, {Krijt}, {Long}, {Lambrechts}, {Rosotti}, {Herczeg}, {Salyk}, {Zhang}, {Bergin}, {Ballering}, {Meyer}, {Bruderer}, \& {Jdiscs Collaboration}}]{Banzatti_2023}
{Banzatti}, A., {Pontoppidan}, K.~M., {Carr}, J.~S., {et~al.} 2023, \apjl, 957, L22, \dodoi{10.3847/2041-8213/acf5ec}

\bibitem[{{Ben{\'\i}tez-Llambay} {et~al.}(2019){Ben{\'\i}tez-Llambay}, {Krapp}, \& {Pessah}}]{fargo_multifluid}
{Ben{\'\i}tez-Llambay}, P., {Krapp}, L., \& {Pessah}, M.~E. 2019, \apjs, 241, 25, \dodoi{10.3847/1538-4365/ab0a0e}

\bibitem[{{Ben{\'\i}tez-Llambay} \& {Masset}(2016)}]{fargo3d}
{Ben{\'\i}tez-Llambay}, P., \& {Masset}, F.~S. 2016, \apjs, 223, 11, \dodoi{10.3847/0067-0049/223/1/11}

\bibitem[{{Bi} {et~al.}(2021){Bi}, {Lin}, \& {Dong}}]{bi2021}
{Bi}, J., {Lin}, M.-K., \& {Dong}, R. 2021, \apj, 912, 107, \dodoi{10.3847/1538-4357/abef6b}

\bibitem[{{Binkert} {et~al.}(2021){Binkert}, {Szul{\'a}gyi}, \& {Birnstiel}}]{binkert2021}
{Binkert}, F., {Szul{\'a}gyi}, J., \& {Birnstiel}, T. 2021, \mnras, 506, 5969, \dodoi{10.1093/mnras/stab2075}

\bibitem[{{Birnstiel} {et~al.}(2012){Birnstiel}, {Klahr}, \& {Ercolano}}]{birnstiel2012}
{Birnstiel}, T., {Klahr}, H., \& {Ercolano}, B. 2012, \aap, 539, A148, \dodoi{10.1051/0004-6361/201118136}

\bibitem[{{Bitsch} {et~al.}(2018){Bitsch}, {Morbidelli}, {Johansen}, {Lega}, {Lambrechts}, \& {Crida}}]{Bitsch_2018}
{Bitsch}, B., {Morbidelli}, A., {Johansen}, A., {et~al.} 2018, \aap, 612, A30, \dodoi{10.1051/0004-6361/201731931}

\bibitem[{{Blandford} \& {Payne}(1982)}]{Blandford_Payne_1982}
{Blandford}, R.~D., \& {Payne}, D.~G. 1982, \mnras, 199, 883, \dodoi{10.1093/mnras/199.4.883}

\bibitem[{{Bodenheimer} \& {Pollack}(1986)}]{Bodenheimer_Pollack_1986}
{Bodenheimer}, P., \& {Pollack}, J.~B. 1986, \icarus, 67, 391, \dodoi{10.1016/0019-1035(86)90122-3}

\bibitem[{Carrera {et~al.}(2021)Carrera, Simon, Li, Kretke, \& Klahr}]{Carrera_2021}
Carrera, D., Simon, J.~B., Li, R., Kretke, K.~A., \& Klahr, H. 2021, The Astronomical Journal, 161, 96, \dodoi{10.3847/1538-3881/abd4d9}

\bibitem[{{Curone} {et~al.}(2025){Curone}, {Facchini}, {Andrews}, {Testi}, {Benisty}, {Czekala}, {Huang}, {Ilee}, {Isella}, {Lodato}, {Loomis}, {Stadler}, {Winter}, {Bae}, {Barraza-Alfaro}, {Cataldi}, {Cuello}, {Fasano}, {Flock}, {Fukagawa}, {Galloway-Sprietsma}, {Garg}, {Hall}, {Izquierdo}, {Kanagawa}, {Lesur}, {Longarini}, {Menard}, {Orihara}, {Pinte}, {Price}, {Rosotti}, {Teague}, {Wafflard-Fernandez}, {Wilner}, {W{\"o}lfer}, {Yen}, {Yoshida}, \& {Zawadzki}}]{curone2025}
{Curone}, P., {Facchini}, S., {Andrews}, S.~M., {et~al.} 2025, \apjl, 984, L9, \dodoi{10.3847/2041-8213/adc438}

\bibitem[{{de Val-Borro} {et~al.}(2006){de Val-Borro}, {Edgar}, {Artymowicz}, {Ciecielag}, {Cresswell}, {D'Angelo}, {Delgado-Donate}, {Dirksen}, {Fromang}, {Gawryszczak}, {Klahr}, {Kley}, {Lyra}, {Masset}, {Mellema}, {Nelson}, {Paardekooper}, {Peplinski}, {Pierens}, {Plewa}, {Rice}, {Sch{\"a}fer}, \& {Speith}}]{deValBorro_2006}
{de Val-Borro}, M., {Edgar}, R.~G., {Artymowicz}, P., {et~al.} 2006, \mnras, 370, 529, \dodoi{10.1111/j.1365-2966.2006.10488.x}

\bibitem[{{Dr{\c{a}}{\.z}kowska} {et~al.}(2019){Dr{\c{a}}{\.z}kowska}, {Li}, {Birnstiel}, {Stammler}, \& {Li}}]{Drazkowska_2019}
{Dr{\c{a}}{\.z}kowska}, J., {Li}, S., {Birnstiel}, T., {Stammler}, S.~M., \& {Li}, H. 2019, \apj, 885, 91, \dodoi{10.3847/1538-4357/ab46b7}

\bibitem[{{Duffell}(2020)}]{duffell_2020}
{Duffell}, P.~C. 2020, \apj, 889, 16, \dodoi{10.3847/1538-4357/ab5b0f}

\bibitem[{{Flaherty} {et~al.}(2020){Flaherty}, {Hughes}, {Simon}, {Qi}, {Bai}, {Bulatek}, {Andrews}, {Wilner}, \& {K{\'o}sp{\'a}l}}]{Flaherty_2020}
{Flaherty}, K., {Hughes}, A.~M., {Simon}, J.~B., {et~al.} 2020, \apj, 895, 109, \dodoi{10.3847/1538-4357/ab8cc5}

\bibitem[{{Flaherty} {et~al.}(2015){Flaherty}, {Hughes}, {Rosenfeld}, {Andrews}, {Chiang}, {Simon}, {Kerzner}, \& {Wilner}}]{Flaherty_2015}
{Flaherty}, K.~M., {Hughes}, A.~M., {Rosenfeld}, K.~A., {et~al.} 2015, \apj, 813, 99, \dodoi{10.1088/0004-637X/813/2/99}

\bibitem[{{Flaherty} {et~al.}(2018){Flaherty}, {Hughes}, {Teague}, {Simon}, {Andrews}, \& {Wilner}}]{Flaherty_2018}
{Flaherty}, K.~M., {Hughes}, A.~M., {Teague}, R., {et~al.} 2018, \apj, 856, 117, \dodoi{10.3847/1538-4357/aab615}

\bibitem[{{Flaherty} {et~al.}(2017){Flaherty}, {Hughes}, {Rose}, {Simon}, {Qi}, {Andrews}, {K{\'o}sp{\'a}l}, {Wilner}, {Chiang}, {Armitage}, \& {Bai}}]{Flaherty_2017}
{Flaherty}, K.~M., {Hughes}, A.~M., {Rose}, S.~C., {et~al.} 2017, \apj, 843, 150, \dodoi{10.3847/1538-4357/aa79f9}

\bibitem[{{Gammie}(1996)}]{gammie1996}
{Gammie}, C.~F. 1996, \apj, 457, 355, \dodoi{10.1086/176735}

\bibitem[{{Gasman} {et~al.}(2025){Gasman}, {Temmink}, {van Dishoeck}, {Kurtovic}, {Grant}, {Sellek}, {Tabone}, {Henning}, {Kamp}, {G{\"u}del}, {Barrado}, {Caratti o Garatti}, {Glauser}, {Waters}, {Arabhavi}, {Jang}, {Kanwar}, {Lienert}, {Perotti}, {Schwarz}, \& {Vlasblom}}]{Gasman_2025}
{Gasman}, D., {Temmink}, M., {van Dishoeck}, E.~F., {et~al.} 2025, \aap, 694, A147, \dodoi{10.1051/0004-6361/202452152}

\bibitem[{{Gressel} {et~al.}(2015){Gressel}, {Turner}, {Nelson}, \& {McNally}}]{Gressel_2015}
{Gressel}, O., {Turner}, N.~J., {Nelson}, R.~P., \& {McNally}, C.~P. 2015, \apj, 801, 84, \dodoi{10.1088/0004-637X/801/2/84}

\bibitem[{Harris {et~al.}(2020)Harris, Millman, van~der Walt, Gommers, Virtanen, Cournapeau, Wieser, Taylor, Berg, Smith, Kern, Picus, Hoyer, van Kerkwijk, Brett, Haldane, del R{\'{i}}o, Wiebe, Peterson, G{\'{e}}rard-Marchant, Sheppard, Reddy, Weckesser, Abbasi, Gohlke, \& Oliphant}]{harris2020array}
Harris, C.~R., Millman, K.~J., van~der Walt, S.~J., {et~al.} 2020, Nature, 585, 357, \dodoi{10.1038/s41586-020-2649-2}

\bibitem[{{Houge} {et~al.}(2025){Houge}, {Krijt}, {Banzatti}, {Blake}, {Pinilla}, {Pontoppidan}, {Trapman}, {Williams}, \& {Zhang}}]{Houge_2025}
{Houge}, A., {Krijt}, S., {Banzatti}, A., {et~al.} 2025, \mnras, 537, 691, \dodoi{10.1093/mnras/staf057}

\bibitem[{Hu {et~al.}(2024)Hu, Li, Bae, \& Zhu}]{hu_2024}
Hu, X., Li, Z.-Y., Bae, J., \& Zhu, Z. 2024, 3D Gap Opening in Non-Ideal MHD Protoplanetary Disks: Asymmetric Accretion, Meridional Vortices, and Observational Signatures.
\newblock \doarXiv{2403.18292}

\bibitem[{{Hu} {et~al.}(2022){Hu}, {Li}, {Zhu}, \& {Yang}}]{hu2022}
{Hu}, X., {Li}, Z.-Y., {Zhu}, Z., \& {Yang}, C.-C. 2022, \mnras, 516, 2006, \dodoi{10.1093/mnras/stac1799}

\bibitem[{Huang {et~al.}(2025)Huang, Yu, Lee, Dong, \& Bai}]{huang2025leakydusttrapsplanetembedded}
Huang, P., Yu, F., Lee, E.~J., Dong, R., \& Bai, X.-N. 2025, Leaky Dust Traps in Planet-Embedded Protoplanetary Disks.
\newblock \doarXiv{2503.19026}

\bibitem[{Hunter(2007)}]{Hunter:2007}
Hunter, J.~D. 2007, Computing in Science \& Engineering, 9, 90, \dodoi{10.1109/MCSE.2007.55}

\bibitem[{Isella {et~al.}(2019)Isella, Benisty, Teague, Bae, Keppler, Facchini, \& Pérez}]{Isella_2019}
Isella, A., Benisty, M., Teague, R., {et~al.} 2019, The Astrophysical Journal Letters, 879, L25, \dodoi{10.3847/2041-8213/ab2a12}

\bibitem[{{Johansen} \& {Youdin}(2007)}]{Johansen_Youdin_2007}
{Johansen}, A., \& {Youdin}, A. 2007, \apj, 662, 627, \dodoi{10.1086/516730}

\bibitem[{Kalyaan {et~al.}(2023)Kalyaan, Pinilla, Krijt, Banzatti, Rosotti, Mulders, Lambrechts, Long, \& Herczeg}]{kalyaan2023effectdustevolutiontraps}
Kalyaan, A., Pinilla, P., Krijt, S., {et~al.} 2023, The Effect of Dust Evolution and Traps on Inner Disk Water Enrichment.
\newblock \doarXiv{2307.01789}

\bibitem[{{Kanagawa} {et~al.}(2015){Kanagawa}, {Muto}, {Tanaka}, {Tanigawa}, {Takeuchi}, {Tsukagoshi}, \& {Momose}}]{Kanagawa_2015}
{Kanagawa}, K.~D., {Muto}, T., {Tanaka}, H., {et~al.} 2015, \apjl, 806, L15, \dodoi{10.1088/2041-8205/806/1/L15}

\bibitem[{{Kimmig} {et~al.}(2020){Kimmig}, {Dullemond}, \& {Kley}}]{Kimming_2020}
{Kimmig}, C.~N., {Dullemond}, C.~P., \& {Kley}, W. 2020, \aap, 633, A4, \dodoi{10.1051/0004-6361/201936412}

\bibitem[{{Kleine} {et~al.}(2020){Kleine}, {Budde}, {Burkhardt}, {Kruijer}, {Worsham}, {Morbidelli}, \& {Nimmo}}]{Kleine_2020}
{Kleine}, T., {Budde}, G., {Burkhardt}, C., {et~al.} 2020, \ssr, 216, 55, \dodoi{10.1007/s11214-020-00675-w}

\bibitem[{Kruijer {et~al.}(2017)Kruijer, Burkhardt, Budde, \& Kleine}]{Kruijer_2017}
Kruijer, T.~S., Burkhardt, C., Budde, G., \& Kleine, T. 2017, Proceedings of the National Academy of Sciences, 114, 6712, \dodoi{10.1073/pnas.1704461114}

\bibitem[{{Kruijer} {et~al.}(2020){Kruijer}, {Kleine}, \& {Borg}}]{Kruijer_2020}
{Kruijer}, T.~S., {Kleine}, T., \& {Borg}, L.~E. 2020, Nature Astronomy, 4, 32, \dodoi{10.1038/s41550-019-0959-9}

\bibitem[{{Lesur} {et~al.}(2014){Lesur}, {Kunz}, \& {Fromang}}]{Lesur_2014}
{Lesur}, G., {Kunz}, M.~W., \& {Fromang}, S. 2014, \aap, 566, A56, \dodoi{10.1051/0004-6361/201423660}

\bibitem[{{Lesur} {et~al.}(2023){Lesur}, {Flock}, {Ercolano}, {Lin}, {Yang}, {Barranco}, {Benitez-Llambay}, {Goodman}, {Johansen}, {Klahr}, {Laibe}, {Lyra}, {Marcus}, {Nelson}, {Squire}, {Simon}, {Turner}, {Umurhan}, \& {Youdin}}]{Lesur_2023}
{Lesur}, G., {Flock}, M., {Ercolano}, B., {et~al.} 2023, in Astronomical Society of the Pacific Conference Series, Vol. 534, Protostars and Planets VII, ed. S.~{Inutsuka}, Y.~{Aikawa}, T.~{Muto}, K.~{Tomida}, \& M.~{Tamura}, 465

\bibitem[{Lubow {et~al.}(1999)Lubow, Seibert, \& Artymowicz}]{Lubow_1999}
Lubow, S.~H., Seibert, M., \& Artymowicz, P. 1999, The Astrophysical Journal, 526, 1001–1012, \dodoi{10.1086/308045}

\bibitem[{{Mathis} {et~al.}(1977){Mathis}, {Rumpl}, \& {Nordsieck}}]{MRN}
{Mathis}, J.~S., {Rumpl}, W., \& {Nordsieck}, K.~H. 1977, \apj, 217, 425, \dodoi{10.1086/155591}

\bibitem[{Miotello {et~al.}(2022)Miotello, Kamp, Birnstiel, Cleeves, \& Kataoka}]{miotello2022settingstageplanetformation}
Miotello, A., Kamp, I., Birnstiel, T., Cleeves, L.~I., \& Kataoka, A. 2022, Setting the Stage for Planet Formation: Measurements and Implications of the Fundamental Disk Properties.
\newblock \doarXiv{2203.09818}

\bibitem[{{M{\"u}ller} {et~al.}(2012){M{\"u}ller}, {Kley}, \& {Meru}}]{Muller_2012}
{M{\"u}ller}, T.~W.~A., {Kley}, W., \& {Meru}, F. 2012, \aap, 541, A123, \dodoi{10.1051/0004-6361/201118737}

\bibitem[{{National Academies of Sciences, Engineering, and Medicine}(2023)}]{NAS2020}
{National Academies of Sciences, Engineering, and Medicine}. 2023, Pathways to Discovery in Astronomy and Astrophysics for the 2020s (Washington, DC: The National Academies Press), \dodoi{10.17226/26141}

\bibitem[{pandas~development team(2020)}]{reback2020pandas}
pandas~development team, T. 2020, pandas-dev/pandas: Pandas, latest,  Zenodo, \dodoi{10.5281/zenodo.3509134}

\bibitem[{{Pinilla} {et~al.}(2012){Pinilla}, {Benisty}, \& {Birnstiel}}]{Pinilla_2012}
{Pinilla}, P., {Benisty}, M., \& {Birnstiel}, T. 2012, \aap, 545, A81, \dodoi{10.1051/0004-6361/201219315}

\bibitem[{{Pinte} {et~al.}(2016){Pinte}, {Dent}, {M{\'e}nard}, {Hales}, {Hill}, {Cortes}, \& {de Gregorio-Monsalvo}}]{pinte2016}
{Pinte}, C., {Dent}, W.~R.~F., {M{\'e}nard}, F., {et~al.} 2016, \apj, 816, 25, \dodoi{10.3847/0004-637X/816/1/25}

\bibitem[{{Pringle} \& {Rees}(1972)}]{Pringle_Rees_1972}
{Pringle}, J.~E., \& {Rees}, M.~J. 1972, \aap, 21, 1

\bibitem[{{Rosotti} {et~al.}(2020){Rosotti}, {Teague}, {Dullemond}, {Booth}, \& {Clarke}}]{rosotti2020}
{Rosotti}, G.~P., {Teague}, R., {Dullemond}, C., {Booth}, R.~A., \& {Clarke}, C.~J. 2020, \mnras, 495, 173, \dodoi{10.1093/mnras/staa1170}

\bibitem[{{Safronov}(1972)}]{Safronov_1972}
{Safronov}, V.~S. 1972, {Evolution of the protoplanetary cloud and formation of the earth and planets.} (Jerusalem: Keter Press)

\bibitem[{{Schneider} \& {Bitsch}(2021{\natexlab{a}})}]{Schneider_Bitsch_A}
{Schneider}, A.~D., \& {Bitsch}, B. 2021{\natexlab{a}}, \aap, 654, A72, \dodoi{10.1051/0004-6361/202141096}

\bibitem[{{Schneider} \& {Bitsch}(2021{\natexlab{b}})}]{Schneider_Bitsch_B}
---. 2021{\natexlab{b}}, \aap, 654, A71, \dodoi{10.1051/0004-6361/202039640}

\bibitem[{{Schneider} \& {Bitsch}(2022)}]{Schneider_Bitsch_2022}
---. 2022, {How drifting and evaporating pebbles shape giant planets (Corrigendum)}, Astronomy \& Astrophysics, Volume 659, id.C3, 3 pp., \dodoi{10.1051/0004-6361/202141096e}

\bibitem[{{Shakura} \& {Sunyaev}(1973)}]{Shakura_Sunyaev_1973}
{Shakura}, N.~I., \& {Sunyaev}, R.~A. 1973, \aap, 24, 337

\bibitem[{{Takeuchi} \& {Lin}(2002)}]{Takeuchi_2002}
{Takeuchi}, T., \& {Lin}, D.~N.~C. 2002, \apj, 581, 1344, \dodoi{10.1086/344437}

\bibitem[{{Teague} {et~al.}(2018){Teague}, {Bae}, {Birnstiel}, \& {Bergin}}]{teague2018}
{Teague}, R., {Bae}, J., {Birnstiel}, T., \& {Bergin}, E.~A. 2018, \apj, 868, 113, \dodoi{10.3847/1538-4357/aae836}

\bibitem[{{Teague} {et~al.}(2016){Teague}, {Guilloteau}, {Semenov}, {Henning}, {Dutrey}, {Pi{\'e}tu}, {Birnstiel}, {Chapillon}, {Hollenbach}, \& {Gorti}}]{Teague_2016}
{Teague}, R., {Guilloteau}, S., {Semenov}, D., {et~al.} 2016, \aap, 592, A49, \dodoi{10.1051/0004-6361/201628550}

\bibitem[{{Temmink} {et~al.}(2024){Temmink}, {van Dishoeck}, {Gasman}, {Grant}, {Tabone}, {G{\"u}del}, {Henning}, {Barrado}, {Caratti o Garatti}, {Glauser}, {Kamp}, {Arabhavi}, {Jang}, {Kurtovic}, {Perotti}, {Schwarz}, \& {Vlasblom}}]{Temmink_2024}
{Temmink}, M., {van Dishoeck}, E.~F., {Gasman}, D., {et~al.} 2024, \aap, 689, A330, \dodoi{10.1051/0004-6361/202450355}

\bibitem[{{Villenave} {et~al.}(2020){Villenave}, {M{\'e}nard}, {Dent}, {Duch{\^e}ne}, {Stapelfeldt}, {Benisty}, {Boehler}, {van der Plas}, {Pinte}, {Telkamp}, {Wolff}, {Flores}, {Lesur}, {Louvet}, {Riols}, {Dougados}, {Williams}, \& {Padgett}}]{villenave2020}
{Villenave}, M., {M{\'e}nard}, F., {Dent}, W.~R.~F., {et~al.} 2020, \aap, 642, A164, \dodoi{10.1051/0004-6361/202038087}

\bibitem[{{Villenave} {et~al.}(2022){Villenave}, {Stapelfeldt}, {Duch{\^e}ne}, {M{\'e}nard}, {Lambrechts}, {Sierra}, {Flores}, {Dent}, {Wolff}, {Ribas}, {Benisty}, {Cuello}, \& {Pinte}}]{villenave2022}
{Villenave}, M., {Stapelfeldt}, K.~R., {Duch{\^e}ne}, G., {et~al.} 2022, \apj, 930, 11, \dodoi{10.3847/1538-4357/ac5fae}

\bibitem[{Virtanen {et~al.}(2020)Virtanen, Gommers, Oliphant, Haberland, Reddy, Cournapeau, Burovski, Peterson, Weckesser, Bright, {van der Walt}, Brett, Wilson, Millman, Mayorov, Nelson, Jones, Kern, Larson, Carey, Polat, Feng, Moore, {VanderPlas}, Laxalde, Perktold, Cimrman, Henriksen, Quintero, Harris, Archibald, Ribeiro, Pedregosa, {van Mulbregt}, \& {SciPy 1.0 Contributors}}]{2020SciPy-NMeth}
Virtanen, P., Gommers, R., Oliphant, T.~E., {et~al.} 2020, Nature Methods, 17, 261, \dodoi{10.1038/s41592-019-0686-2}

\bibitem[{{Wafflard-Fernandez} \& {Lesur}(2023)}]{Wafflard_Lesur_2023}
{Wafflard-Fernandez}, G., \& {Lesur}, G. 2023, \aap, 677, A70, \dodoi{10.1051/0004-6361/202245305}

\bibitem[{{Wafflard-Fernandez} \& {Lesur}(2025)}]{Wafflard_Lesur_2025}
---. 2025, \aap, 696, A8, \dodoi{10.1051/0004-6361/202453541}

\bibitem[{{Walsh} {et~al.}(2011){Walsh}, {Morbidelli}, {Raymond}, {O'Brien}, \& {Mandell}}]{walsh2011}
{Walsh}, K.~J., {Morbidelli}, A., {Raymond}, S.~N., {O'Brien}, D.~P., \& {Mandell}, A.~M. 2011, \nat, 475, 206, \dodoi{10.1038/nature10201}

\bibitem[{{Weber} {et~al.}(2018){Weber}, {Ben{\'\i}tez-Llambay}, {Gressel}, {Krapp}, \& {Pessah}}]{Weber_2018}
{Weber}, P., {Ben{\'\i}tez-Llambay}, P., {Gressel}, O., {Krapp}, L., \& {Pessah}, M.~E. 2018, \apj, 854, 153, \dodoi{10.3847/1538-4357/aaab63}

\bibitem[{{Weidenschilling}(1977)}]{weidenschilling1977}
{Weidenschilling}, S.~J. 1977, \apss, 51, 153, \dodoi{10.1007/BF00642464}

\bibitem[{{Whipple}(1972)}]{whipple1972}
{Whipple}, F.~L. 1972, in From Plasma to Planet, ed. A.~{Elvius}, 211

\bibitem[{{Wu} {et~al.}(2023){Wu}, {Chen}, {Jiang}, {Dong}, {Mac{\'\i}as}, {Lin}, {Rosotti}, \& {Elbakyan}}]{Wu_2023}
{Wu}, Y., {Chen}, Y.-X., {Jiang}, H., {et~al.} 2023, \mnras, 523, 2630, \dodoi{10.1093/mnras/stad1553}

\bibitem[{{Youdin} \& {Goodman}(2005)}]{Youdin_Goodman_2005}
{Youdin}, A.~N., \& {Goodman}, J. 2005, \apj, 620, 459, \dodoi{10.1086/426895}

\bibitem[{{Yun} {et~al.}(2019){Yun}, {Kim}, {Bae}, \& {Han}}]{yun2019}
{Yun}, H.~G., {Kim}, W.-T., {Bae}, J., \& {Han}, C. 2019, \apj, 884, 142, \dodoi{10.3847/1538-4357/ab3fab}

\bibitem[{{Zhang} {et~al.}(2018){Zhang}, {Zhu}, {Huang}, {Guzm{\'a}n}, {Andrews}, {Birnstiel}, {Dullemond}, {Carpenter}, {Isella}, {P{\'e}rez}, {Benisty}, {Wilner}, {Baruteau}, {Bai}, \& {Ricci}}]{Zhang_2018}
{Zhang}, S., {Zhu}, Z., {Huang}, J., {et~al.} 2018, \apjl, 869, L47, \dodoi{10.3847/2041-8213/aaf744}

\bibitem[{{Zhu} {et~al.}(2012){Zhu}, {Nelson}, {Dong}, {Espaillat}, \& {Hartmann}}]{Zhu_2012}
{Zhu}, Z., {Nelson}, R.~P., {Dong}, R., {Espaillat}, C., \& {Hartmann}, L. 2012, \apj, 755, 6, \dodoi{10.1088/0004-637X/755/1/6}

\end{thebibliography}
\bibliographystyle{aasjournal}



\end{document}